\documentclass[a4paper,oneside,11pt]{article}
\usepackage[margin=26mm]{geometry}

\usepackage[utf8]{inputenc}
\usepackage[T1]{fontenc}
\usepackage{fancyhdr}
\usepackage{lastpage}
\usepackage{titling}
\usepackage{caption}
\usepackage{authblk}

\usepackage{booktabs}
\usepackage{csquotes}
\usepackage{enumitem}
\usepackage{tabularx}
\usepackage{xcolor}
\usepackage[
	backend=biber,
	safeinputenc,
	bibencoding=utf8,
	sorting=none,
	maxcitenames=10,
	maxbibnames=10,
	style=numeric,
	citestyle=numeric-comp]{biblatex}
\usepackage{graphicx}

\usepackage{hyperref}

\usepackage{graphicx}
\usepackage{textcomp}
\usepackage{amsmath}
\usepackage{amsthm}

\usepackage{tablefootnote}

\usepackage{algorithm}
\usepackage[noend]{algpseudocode}
\usepackage{bm}

\usepackage{tikz}
\usepackage{pgfplots}
\usetikzlibrary{arrows.meta,shapes,positioning}

\usepackage{booktabs}

\newcommand{\dfourquery}{\texttt{d4-query}}

\tikzset{
  dag/.style={
    ->,
    nodes={draw, circle, minimum size = .6cm},
    >=Stealth[],
	level 1/.style={sibling distance=35mm,level distance=1.2cm},
	level 2/.style={sibling distance=15mm,level distance=1cm},
	level 3/.style={sibling distance=15mm,level distance=1cm}
  },
}

\usepackage{listings}

\lstdefinelanguage{ctodformat}{
	keywords = {O,A,L,nnf},
	morecomment=[l]{\#},
	commentstyle=\itshape\color{purple!40!black}
}

\lstdefinelanguage{dfourformat}{
	keywords = {o,a,t,f},
	morecomment=[l]{\#},
	commentstyle=\itshape\color{purple!40!black}
}

\tikzstyle{orstyle} = [inner sep = 2.5pt]
\tikzstyle{andstyle} = []
\tikzstyle{negatedliteralstyle} = [inner sep = 1.5pt]

\newboolean{blind}
\setboolean{blind}{false}

\newboolean{final}
\setboolean{final}{true}
\ifthenelse{\boolean{final}}{
	\newcommand{\todo}[1]{}
	
	\newcommand{\hint}[1]{}
	\newcommand{\review}[1]{}
	\newcommand{\questionForReview}[1]{}
	\newcommand{\saveSpaceHere}[1]{}
	
}{
	\newcommand{\todo}[1]{{\color{red}\textit{[#1]}}}
	
	\newcommand{\hint}[1]{{\color{hint}\textit{[#1]}}}
	\newcommand{\review}[1]{\textcolor{orange}{[Review required: #1]}}
	\newcommand{\questionForReview}[1]{\review{Question for Review: #1}}
	\newcommand{\saveSpaceHere}[1]{\hint{Save space here: #1}}
	
}

\ifthenelse{\boolean{final}}{
	\ifthenelse{\boolean{blind}}{
		\newcommand{\doubleblind}[2]{#2}
	}{
		\newcommand{\doubleblind}[2]{#1}
	}
}{
	\newcommand{\doubleblind}[2]{\textcolor{green}{#1} \textcolor{purple}{#2}}
}

\newcommand{\expone}{data/fm_summary.csv}
\newcommand{\exptwo}{data/features_summary.csv}
\newcommand{\expthree}{data/pc_summary.csv}

\definecolor{colorc2d}{HTML}{cc78bc}
\definecolor{colord4}{HTML}{029e73}
\definecolor{colord4old}{HTML}{0173b2}
\definecolor{colordsharp}{HTML}{d55e00}
\definecolor{colorssat}{HTML}{949494}

\tikzstyle{pointplot} = [only marks,opacity=0.8]

\tikzstyle{dpll} = [pointplot, mark options={draw=colorssat, fill=colorssat!50}]
\tikzstyle{countantom} = [dpll,mark=pentagon*]
\tikzstyle{sharpsat} = [dpll,mark=square*]
\tikzstyle{ganak} = [dpll,mark=triangle*]

\tikzstyle{compiler} = [pointplot, mark=otimes]

\tikzstyle{ctod} = [mark options={draw=colorc2d, fill=colorc2d!50}]
\tikzstyle{dsharp} = [mark options={draw=colordsharp, fill=colordsharp!50}]
\tikzstyle{dfour} = [mark options={draw=colord4, fill=colord4!50}]
\tikzstyle{dfourold} = [mark options={draw=colord4old, fill=colord4old!50}]
\tikzstyle{dfourquery} = [pointplot, mark=o,mark options={draw=colord4, fill=colord4!50}]

\tikzstyle{queryddnnf} = [pointplot,mark=asterisk]
\tikzstyle{ddknnife} = [pointplot,mark=diamond*]

\tikzstyle{optall} = [pointplot,mark=diamond*,mark options={draw=red, fill=red!50}]
\tikzstyle{optrectree} = [pointplot,mark=asterisk]
\tikzstyle{optrecdag} = [pointplot,mark=o]
\tikzstyle{optwofold} = [pointplot,mark=pentagon*]
\tikzstyle{optwocore} = [pointplot,mark=square*]
\tikzstyle{optwomark} = [pointplot,mark=triangle*]

\algnewcommand{\IIf}[1]{\State\algorithmicif\ #1\ \algorithmicthen}
\algnewcommand{\EndIIf}{\unskip\ \algorithmicend\ \algorithmicif}
\algnewcommand{\ElseIIf}[1]{\algorithmicelse\ #1}
\algnewcommand{\IfThenElse}[3]{% \IfThenElse{<if>}{<then>}{<else>}
	\State \algorithmicif\ #1\ \algorithmicthen\ #2\ \algorithmicelse\ #3}

\definecolor{colorcardfeature}{RGB}{0,114,178}
\definecolor{colorcardpc}{RGB}{213,94,0}

\newcommand{\countantom}{\texttt{countAntom}}

\newcommand{\sharpsat}{\texttt{sharpSAT}}

\newcommand{\ganak}{\texttt{Ganak}}

\newcommand{\ctod}{\texttt{c2d}}
\newcommand{\dsharp}{\texttt{dSharp}}
\newcommand{\dfour}{\texttt{d4}}

\newcommand{\ddknnife}{\texttt{ddnnife}}
\newcommand{\queryddnnf}{\texttt{query-ddnnf}}

\newcommand{\ssat}{\texttt{\#}SAT}

\usepackage{xpunctuate}

\providecommand{\eg}[0]{e.g\xperiod}
\providecommand{\ie}[0]{i.e\xperiod}
\providecommand{\etal}[0]{et~al\xperiod}
\providecommand{\cf}[0]{cf\xperiod}

\providecommand{\wrt}[0]{w.r.t\xperiod}

\theoremstyle{definition}
\newtheorem{definition}{Definition}

\usepackage{enumitem}
\usepackage{calc}
\newcommand*{\researchquestion}[1]{\item[\textbf{RQ#1}]}
\newcommand*{\researchquestionlabelwidthof}[1]{\labelsep+\widthof{\textbf{RQ#1}}}

\newcommand{\latexutilautorefoverride}[2]{
	\ifcsname #1\endcsname
		\expandafter\renewcommand\csname #1\endcsname{#2}
	\else
		\expandafter\newcommand\csname #1\endcsname{#2}
	\fi	
}

\latexutilautorefoverride{figureautorefname}{Figure}
\latexutilautorefoverride{equationautorefname}{Equation}
\latexutilautorefoverride{sectionautorefname}{Section}
\latexutilautorefoverride{subsectionautorefname}{Section}
\latexutilautorefoverride{subsubsectionautorefname}{Section}
\latexutilautorefoverride{definitionautorefname}{Definition}
\latexutilautorefoverride{footnoteautorefname}{Footnote}
\latexutilautorefoverride{algorithmautorefname}{Algorithm}

\begin{document}

\author[1]{Chico Sundermann}
\author[1]{Heiko Raab}
\author[1]{Tobias Heß}
\author[1]{Thomas Thüm}
\author[2]{Ina Schaefer}

\affil[1]{University of Ulm, Germany}
\affil[2]{Karlsruhe Institute of Technology, Germany}

\title{Exploiting d-DNNFs for Repetitive Counting Queries on Feature Models}

\maketitle
\thispagestyle{empty}

\begin{abstract}
\noindent
Feature models are commonly used to specify the valid configurations of a product line.
In industry, feature models are often complex due to a large number of features and constraints.
Thus, a multitude of automated analyses have been proposed.
Many of those rely on computing the number of valid configurations which typically depends on solving a \ssat{} problem, a computationally expensive operation.
Further, most counting-based analyses require numerous \ssat{} computations on the same feature model.
In particular, many analyses depend on multiple computations for evaluating the number of valid configurations that include certain features or conform to partial configurations.
Instead of using expensive repetitive computations on highly similar formulas, we aim to improve the performance by reusing knowledge between these computations.
In this work, we are the first to propose reusing d-DNNFs for performing efficient repetitive queries on features and partial configurations.
Our empirical evaluation shows that our approach is up-to 8,300 times faster (99.99\% CPU-time saved) than the state of the art of repetitively invoking \ssat{} solvers.
Applying our tool \ddknnife{} reduces runtimes from days to minutes compared to using \ssat{} solvers.
\end{abstract}

\section{Introduction}

A product line describes a family of related products which share a set of features~\cite{BSRC10,LSR07,M10,B05,TAK+:CSUR14}.
Typically, not every configuration of these features is valid (\ie, represents an actual product).
Feature models are the de facto standard for specifying the set of valid software configurations~\cite{B05,BSRC10,PBL05} and are also frequently used in other domains~\cite{FAR:SPLC20,HPS+:GPCE22,HST+:MODELS22,KTS+:ESECFSE17}.
In general, a feature model consists of the underlying features and a set of constraints describing dependencies between features.
Due to high numbers of features and constraints, it is typically infeasible to analyze feature models manually~\cite{BSRC10}.
Hence, automated analyses, such as checking whether a given configuration is valid, are demanded~\cite{BTR:SEKE05,MTS+17,BSRC10,SSK+:VaMoS20,BSTRC:VaMoS07,ACLF:SCP13,KTS+:ICSE18}.
% However, with current solutions for model counting being very computationally expensive~\cite{PLP:ASE11,KZK:LoCoCo10,SHN+:EMSE23}, analyses depending on model counting have been only sparsely used~\cite{HST+:MODELS22,ACLF:SCP13,KPK+:SPLC17,BSTRC:VaMoS07}.  

Numerous practice-relevant analyses depend on computing the number of valid configurations for a feature model~\cite{SNB+:VaMoS21,HFCA:IJSEKE13,FHCC:TSE14,KZK:LoCoCo10,CE:SPLC11,HFCC:IET11}.
To employ standardized solvers, feature models are often translated to propositional logic for analysis~\cite{CW:SPLC07,MWC:SPLC09}.
While SAT solvers are used for checking the satisfiability, \ssat{} solvers are dedicated to compute the number of satisfying assignments~\cite{STS:VaMoS20,KZK:LoCoCo10}.
Even though \ssat{} is computationally complex~\cite{V:SICOMP79,J:A92,T:SAT06,FHH:JEA21}, \doubleblind{in recent work~\cite{STS:VaMoS20} we observed}{recent empirical results~\cite{STS:VaMoS20} indicate} that state-of-the-art \ssat{} solvers can compute the number of valid configurations for many real-world feature models within a few seconds.

\doubleblind{The majority of applications we identified in previous work~\cite{SNB+:VaMoS21} require numerous \ssat{} invocations.}{However, for the majority of applications based on feature-model counting, a single \ssat{} invocation is not sufficient~\cite{SNB+:VaMoS21}.}
For example, it can be beneficial to prioritize features that appear in many valid configurations for resource allocation.
Using an off-the-shelf \ssat{} solver as black box, it may be necessary to invoke a separate \ssat{} computation for every single feature~\cite{SNB+:VaMoS21,HFCA:IJSEKE13}.
While several incremental solvers (\ie, that reuse information between queries) for regular SAT were considered in the literature~\cite{NRS:SAT14,ES:ENTCS03}, incremental \ssat{} was not considered yet.
As feature models with up to tens of thousands of features have been reported~\cite{KTS+:ESECFSE17,STS:VaMoS20}, repetitive counting queries without reusing information may require days of computation time~\cite{STS:VaMoS20}.
While such runtimes can be remedied to some degree (\eg, by using parallelization), feature-model analyses are often applied in interactive applications~\cite{MTS+17,ACLF:SCP13,BSTRC:VaMoS07,FAR:SPLC20,KTS+:ICSE18} or continuous-integration environments~\cite{PKR+:VaMoS21,LMVA:SPLC20}, mandating substantially shorter runtimes and limited resource usage.
In particular, these requirements arose from collaborations with the automotive industry, where we integrated our prototype to reduce resource consumption.

With ever-growing feature models~\cite{STS:VaMoS20,IF:JSS10,LSB+:SPLC10}, one is therefore interested in reducing runtime required for analyses.
We propose to employ knowledge compilation, which we use to reduce the computational effort when performing multiple queries~\cite{DM:JAIR02,D:AAAI02,MMBH:AAAI10,LM:IJCAI17}.
Hereby, the original formula is translated once to a target format in an offline phase. Afterwards, the resulting format can be used for online computations (\eg, for \ssat{}) with typically polynomial-time complexity with respect to the size of the knowledge-compilation artifact~\cite{DM:JAIR02}.

The deterministic decomposable negation normal form (d\nobreakdash-DNNF) enables computing the number of satisfying assignments in linear time with respect to its number of nodes~\cite{D:JACM01}.
Compilation to d-DNNF is similarly complex as solving a \ssat{} problem as the d-DNNF is constructed from the trace of one \ssat{} invocation with available tools~\cite{D:AAAI02,LM:IJCAI17,MMBH:AAAI10}.
\doubleblind{In our recent work~\cite{STS:VaMoS20}, we}{Recent work~\cite{STS:VaMoS20}} showed that d\nobreakdash-DNNF compilers~\cite{D:AAAI02,OD:CP14,LM:IJCAI17,HD:IJCAI05,MMBH:AAAI10} are generally faster for feature models than other knowledge compilers that enable model counting.
For instance, compiling to d-DNNF is substantially faster than compiling to binary decision diagrams~\cite{SHN+:EMSE23,HST:SPLC21}.
Thus, d-DNNFs are promising for feature-model counting.

While the compilation to d-DNNF is well researched~\cite{D:AAAI02,OD:CP14,LM:IJCAI17,HD:IJCAI05,MMBH:AAAI10}, reusing a given d\nobreakdash-DNNF is only explored sparsely~\cite{D:ANCL01,SGRM:LPAIR18}.
In particular, reusing a d-DNNF for multiple counting queries has not been evaluated before in software engineering or any other domain.
In related work, Sharma~\etal~\cite{SGRM:LPAIR18} traverse d-DNNFs to derive satisfying assignments for uniform random sampling.
Also, some publications used d-DNNF compilers for counting as black-box \ssat{} solvers without reusing the d-DNNF, essentially compiling a new d-DNNF for every query~\cite{STS:VaMoS20,KZK:LoCoCo10}.

In this work, we propose to \textit{reuse} d-DNNFs for repetitive counting queries which we apply for computing number of valid configurations that (1) contain a certain feature or (2) conform to a partial configuration (\ie, include and/or exclude sets of features).
We showcase the vast improvement of our approach over the state-of-the-art with a large-scale empirical evaluation.
In particular, we compile a given feature model to d-DNNF with a one time effort using existing compilers~\cite{D:AAAI02,LM:IJCAI17,MMBH:AAAI10} and then reuse the d-DNNF for numerous counting-based computations.
Thereby, our work focuses on the efficient reuse of d-DNNFs with the following contributions.

First, we present algorithms and several novel optimizations for repetitive counting queries on a given d-DNNF (\cf \autoref{sec:concept:queries} and \autoref{sec:concept:optimizations}).
Some of the basics for the presented algorithms are inspired by ideas from Darwiche~\cite{D:ANCL01}.
Still, Darwiche's idea was not followed up further with an implementation or empirical evaluation yet.
Further, we contribute a set of novel optimizations which vastly reduce the number of required traversals through the d-DNNF and also accelerate single traversals.

Second, we provide an open-source tool \ddknnife{}\footnote{\url{https://github.com/SoftVarE-Group/d-dnnf-reasoner/releases/tag/March2023}} including all presented algorithms and optimizations (\cf \autoref{sec:toolsupport}).
Our d-DNNF reasoner \ddknnife{} takes a d-DNNF as input supporting all formats given by the the popular compilers~\cite{D:AAAI02,MMBH:AAAI10,LM:IJCAI17}.
Given the d-DNNF, \ddknnife{} allows the user to perform counting queries on the d-DNNF.

Third, we perform a large-scale empirical evaluation comparing the reuse of d-DNNFs with state-of-the-art \ssat{} solvers on 62 industrial feature models (\cf \autoref{sec:evaluation}).
Hereby, we compare our tool \ddknnife{} paired with each available d-DNNF compiler to state-of-the-art solutions.
The required data, tools, and framework are publicly available for replicability and reproducibility.\footnote{\url{https://github.com/SoftVarE-Group/exploiting-ddnnfs-eval/releases/tag/March2023}}

\section{Motivating Example}
\begin{figure}
	\centering
	\includegraphics[width=0.65\textwidth]{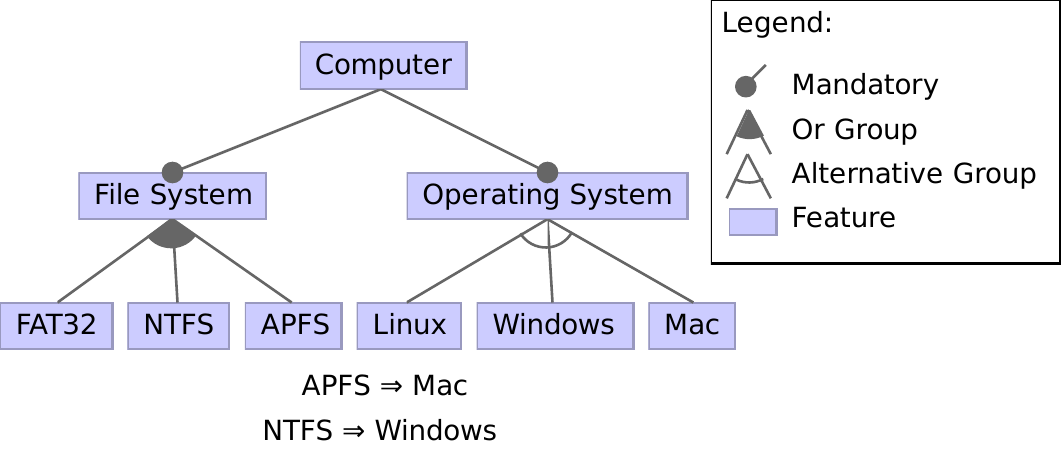}
	\caption{Feature Model Adapted from Krieter~\etal~\cite{KTS+:ICSE18}}
	\label{fig:motivation:pc}
\end{figure}

In this section, we showcase use cases of analyses dependent on multiple counting queries on a small example.
\autoref{fig:motivation:pc} shows the feature model representing a simplified PC.
Each PC requires at least one \texttt{File System} (denoted as or group) and exactly one \texttt{Operating System} (denoted as alternative group).
Furthermore, using \texttt{APFS} or \texttt{NTFS} requires using \texttt{Mac} or \texttt{Windows}, respectively, denoted by the cross-tree constraints below the tree.
In this model, \texttt{Linux} can be selected in a single configuration with \texttt{FAT32} as \texttt{File System}.
\texttt{Windows} can be combined with \texttt{FAT32}, \texttt{NTFS}, or both.
Analogously, \texttt{Mac} appears in three valid configurations including combinations of \texttt{FAT32} and \texttt{APFS}.
Overall, there are seven valid configurations.

Computing the number of valid configurations enables various analyses.
\doubleblind{In our recent work~\cite{SNB+:VaMoS21}, we}{Sundermann~\etal~\cite{SNB+:VaMoS21}}  presented applications dependent on computing the number of valid configurations (1) collected from the literature~\cite{HFCC:IET11,HFME:ICSE19,HFCA:IJSEKE13,KZK:LoCoCo10,BG:SQJ11,CK05} and (2) inspired from industry collaborations.
In that recent work, \doubleblind{we}{the authors} identified 21 applications that are relevant for detecting errors and smells, sampling, maintainability, supporting development, economical estimations, and user guidance.
14 of the 21 applications depend on multiple \ssat{} computations for the same feature model.

As an example, the number of valid configurations containing a specific feature can be used to quantify the impact of an error~\cite{KZK:LoCoCo10,SNB+:VaMoS21}.
It may be sensible to address an issue that appears in more valid configurations before an issue appearing in only a fraction of valid configurations.
For instance, an error in \texttt{FAT32} has an impact on five valid configurations while an error in \texttt{NTFS} impacts only two valid configurations.

Feature-model counting can also be used during the \textit{development process} of a product line.
For instance, it may be beneficial to prioritize features that allow building more distinct valid configurations~\cite{SNB+:VaMoS21}.
Consider the following scenario.
For the PC product line shown in \autoref{fig:motivation:pc}, each operating system is already implemented and the developers have to decide which \texttt{File System} to implement first.
For example, after implementing \texttt{FAT32}, we can build each valid configuration that includes \texttt{FAT32} and excludes \texttt{NTFS} and \texttt{APFS}.
Computing the number of valid configurations that include and exclude respective features shows that implementing \texttt{FAT32} induces three valid configurations while \texttt{NTFS} and \texttt{APFS} allow only one.

While both presented example analyses require only few counting queries for our running example, such analyses often require numerous queries on large feature models in practice.
For instance, homogeneity~\cite{SNB+:VaMoS21,HFME:ICSE19,FHCC:TSE14}, which describes the similarity of configurations induced by the feature model, requires a counting query for every single feature.
Especially, when dealing with large industrial feature models containing thousands of features, performing such analyses may induce large runtimes.
For each of the 14 applications requiring multiple \ssat{} invocations, it may be beneficial to reuse d-DNNFs over multiple computations.

\section{Background} \label{sec:background}
In this section, we provide the background required for the following sections.
First, we introduce feature models and feature-model counting formally.
Second, we introduce d-DNNFs and how to use their structure for model counting.

\subsection{Feature-Model Counting} \label{sec:featuremodelcounting}
For a given product line, a feature model is typically used to specify the set of valid configurations~\cite{BSRC10}.
Formally, we define feature models as tuple $\mathit{FM} = (\mathit{FT}, \mathit{CT})$ with a set of features $\mathit{FT}$ and constraints $\mathit{CT}$ between these features.
The set of constraints $\mathit{CT}$ includes both hierarchical constraints that describe the parent-child relationship, and cross-tree constraints.
For the sake of simplicity, we consider the constraints to be a conjunction of \textit{propositional} formulas (\ie, each constraint in $\mathit{CT}$ has to be satisfied to satisfy the feature model).
We refer to the resulting formula as $\mathit{F_{FM}} = \bigwedge_{c \in \mathit{CT}} c$.
It is well-known that each hierarchical constraint (\ie, alternative, or, mandatory, and optional) can be translated to a propositional formula~\cite{BSRC10,CW:SPLC07}.

A configuration $C = (I,E)$ consists of two sets of features which describe features included ($I \subseteq \mathit{FT}$)  in the configuration and excluded ($E \subseteq \mathit{FT}$) in the configuration.
A feature cannot be included and excluded in the same configuration (\ie, $I \cap E = \emptyset$).
If all features are included or excluded (\ie, $I \cup E = \mathit{FT}$), the configuration is complete.
Otherwise, the configuration is partial~\cite{BSRC10}.
We consider a configuration $C = (I, E)$ to be valid, iff it is complete and satisfies all constraints $\mathit{CT}$ imposed by the feature model~\cite{BSRC10}.
We refer to the set of all valid configurations induced by $\mathit{FM}$ as $\mathit{VC_{FM}}$.

Feature-model counting refers to computing the number of valid configurations for a given feature model.
\doubleblind{In previous work~\cite{SNB+:VaMoS21}, we introduce}{Sundermann~\etal~\cite{SNB+:VaMoS21} introduce} three types of feature-model counting that compute the number of valid configurations (1) of the entire feature model, (2) that include a certain feature, and (3) that are induced by a partial configuration (\ie, include some and exclude other features). These three types cover all the 21 applications that we identified.

\noindent\begin{definition}[Cardinality of a Feature Model]
	The cardinality $\texttt{\#}\mathit{FM}$ of a feature model $\mathit{FM} = (\mathit{FT},\mathit{CT})$ corresponds to the cardinality of the set of valid configurations $\mathit{VC_{FM}}$ (\ie,  $\texttt{\#}\mathit{FM} = |\mathit{VC_{FM}}|$).
\end{definition}

\begin{definition}[Cardinality of a Feature]
	The cardinality $\texttt{\#}f$ of a feature $f$ in a feature model $\mathit{FM} = (\mathit{FT}, \mathit{CT})$ corresponds to the cardinality of the set of valid configurations that include $f$ (\ie, $\texttt{\#}f = ~|~ \{C = (I,E) \, | \, C \in \mathit{VC_{FM}}, f \in I\}|$).
\end{definition}

\begin{definition}[Cardinality of a Partial Configuration]
	The cardinality $\#C_p$ of a partial configuration $C_p = (I_p, E_p)$ given a feature model $\mathit{FM} = (\mathit{FT}, \mathit{CT})$ corresponds to the cardinality of the set of valid configurations that include each feature $i \in I_p$ and exclude each feature $e \in E_p$ (\ie, $\texttt{\#}C_p = \, \left| \, \{C = (I, E) \, | \, C \in \mathit{VC_{FM}}, I_p \subseteq I, E_p \subseteq E \}\right|$).
\end{definition}

Many feature-model analyses are reduced to solving SAT or \ssat{} problems on a propositional formula representing the feature model~\cite{HFCA:IJSEKE13,ZZM04,BSRC10,KZK:LoCoCo10,TBK:ICSE09,MWC:SPLC09,LGCR:SPLC15}.
For example, a feature model $\mathit{FM} = (\mathit{FT}, \mathit{CT})$ is void (\ie, induces no valid configuration~\cite{BSRC10}) iff the formula $\mathit{F_{FM}}$ is not satisfiable.
The three types of cardinality defined above can be reduced to \ssat{} invocations~\cite{SNB+:VaMoS21,HFCA:IJSEKE13}. The cardinality of a feature model $\#\mathit{FM}$ is equivalent to the number of satisfying assignments of $F_{FM}$ (\ie, \ssat{}($\mathit{F_{FM}}$)).
The cardinality of a feature $\#f$ can be computed via \ssat{}$(\mathit{F_{FM}} \wedge f)$. Analogously, the cardinality $\#C_p$ of a partial configuration $C_p = (I,E)$  can be reduced to \ssat{}$(\mathit{F_{FM}} ~\land~ \bigwedge_{i \in I} i ~\land~ \bigwedge_{e \in E} \neg e)$.

\subsection{Model Counting with d-DNNFs} \label{sec:background:ddnnf}
Typically, tools based on propositional logic, such as SAT and \ssat{} solvers, rely on normal forms as representation for input formulas. The most commonly used format is conjunctive normal form (CNF)~\cite{T:SAT06,BSB:SAT15,SRSM:IJCAI19,D:AAAI02,D:ECAI04,LM:IJCAI17,MMBH:AAAI10,B:JSAT08,BP:AAAI00,KMM:QEST13,KLMT:IJCAI13,OD:IJCAI15,SBB+:SAT04}.
However, CNFs do not allow model counting in polynomial time~\cite{DM:JAIR02,V:SICOMP79}.

In contrast, the \textit{d-DNNF} (\textit{d}eterministic \textit{D}ecomposable \textit{N}egation \textit{N}ormal \textit{F}orm) format supports linear-time model counting with respect to the number of d-DNNF nodes~\cite{DM:JAIR02,D:AAAI02,MMBH:AAAI10}.
The language d-DNNF is a subset of negation normal form (NNF) for which determinism and decomposability holds~\cite{D:JACM01}.
Often, d-DNNFs are further restricted to be \textit{smooth} for simplifying and accelerating algorithms on a d-DNNF~\cite{D:ECAI04,DM:JAIR02,D:ANCL01,MMBH:AAAI10}.

\begin{definition}[Negation Normal Form~\cite{D:JACM01}]
	A propositional formula $F$ is in \textit{negation normal form} iff $\land, \lor,$ and $\neg$ are the only logical operators and
	negations ($\neg$) only appear directly in front of variables.
\end{definition}

\autoref{algo:ddnnf:base} shows how to compute the number of satisfying assignments for a given d-DNNF.
Starting from the root, the algorithm traverses the d-DNNF and applies rules depending on the node type (\ie, conjunction, disjunction, and literals) exploiting the properties decomposability, determinism, and smoothness.
In the following, we explain these three properties of d-DNNFs and their impact on \autoref{algo:ddnnf:base}.

\begin{algorithm}[tb]
	\caption{Model Counting via d-DNNF Traversal}
	\label{algo:ddnnf:base}
	\begin{algorithmic}[1]
		\Procedure{\#node}{node}
		\If {node is \texttt{And}}
		\State \textbf{return} $\prod_{c \in \texttt{children}}$ \Call{\#node}{c}
		\ElsIf{node is \texttt{Or}}
		\State \textbf{return} $\sum_{c \in \texttt{children}}$ \Call{\#node}{c}
		\ElsIf{node is \texttt{Literal}}
		\State \textbf{return} 1
		\EndIf
		\EndProcedure
	\end{algorithmic}
\end{algorithm}

\begin{definition}[Decomposability~\cite{D:AAAI02}]
	A propositional formula $F$ is \textit{decomposable} iff for each conjunction $C = C_1 \wedge C_2 \wedge \ldots \wedge C_n$ the sets of variables of each conjunct $C_i$ are disjunct (\ie, $\forall\, i,j: i \neq j: vars(C_i) \cap vars(C_j) = \emptyset$).
\end{definition}

With decomposability, the number of satisfying assignments of a conjunction is the product of its children's numbers of satisfying assignments, as seen in Lines 2-3 of \autoref{algo:ddnnf:base}.
If a formula $F$ is decomposable, the conjuncts of each conjunction share no variables.
Thus, we can combine the satisfying assignments pairwise (\ie, building the cross-product) for each of the conjuncts.
Due to the pairwise combination, the size of the resulting set (\ie, the number of satisfying assignments) is the product of sizes for the respective conjuncts.

Consider the example formula $F = (A \vee B) \wedge (C \vee \neg C)$. The left part $A \vee B$ induces the three satisfying assignments $\{A,B\}, \{A, \neg B\}, \{\neg A , B\}$. The right part $C \vee \neg C$ induces the two satisfying assignments $\{C\}, \{\neg C\}$. Pairwise combining both sets results in $3 \cdot 2 = 6$ satisfying assignments for $F$ (\ie, $\{A,B,C\}$, $\{A,B,\neg C\}$, $\{A, \neg B, C\}$, $\{A, \neg B, \neg C\}$, $\{\neg A, B, C\}$, $\{\neg A, B, \neg C\}$).
Without decomposability, the described procedure may lead to faulty results as the combined assignments may be conflicting or not disjunct leading to faulty results when counting.
% Consider the non-decomposable formula $(A \lor B) \land (B \lor C)$.
% Both conjuncts induce three solutions, namely \{$\{A,B\}$, $\{A, \neg B\}$, $\{\neg A,B\}$\} and $\{\{C,B\}, \{C, \neg B\}$, $\{\neg C,B\}\}$.
% However, due to the shared variables the satisfying assignments of the conjuncts impact each other and cannot be arbitrarily combined (\eg, $\{A, \neg B\}$ and $\{\neg C, B\}$ has a conflict).
% Thus, building the product for these numbers of satisfying assignments wrongfully leads to nine instead of five satisfying assignments.

\begin{definition}[Determinism~\cite{D:AAAI02}]
	A propositional formula $F$ is \textit{deterministic} iff for each disjunction $D = D_1 \vee D_2 \vee \ldots \vee D_m$, the disjuncts share no common satisfying assignments (\ie, $\forall \, i,j: i \neq j: D_i \wedge D_j = \bot$).
\end{definition}

With determinism, the number of satisfying assignments of a disjunction is the sum of its children's numbers as seen in Lines 4--5 of \autoref{algo:ddnnf:base}.
For a disjunction, each assignment that satisfies one of the disjuncts also satisfies the disjunction.
Therefore, the set of satisfying assignments is the union over the set of the conjuncts.
For a deterministic disjunction, there are no solutions shared between disjuncts.
Thus, when conjoining the set of satisfying assignments of all disjuncts there are no duplicate entries.

Consider the example formula $F = (A \wedge B) \vee (\neg B)$. $(A \wedge B)$ and $(\neg B)$ induce $\{A,B\}$ and $\{A,\neg B\}$, $\{\neg A, \neg B\}$ as satisfying assignments, respectively. The set of satisfying assignments for $F$ is the union \{$\{A,B\}$, $\{A,\neg B\}$, $\{\neg A, \neg B\}$\} of both sets. As the sets are disjunct by definition, the number of satisfying assignments is the sum of the number of satisfying assignments for its children.
Without determinism, the described procedure may lead to faulty results as there may be duplicate satisfying assignments wrongfully increasing the number of satisfying assignments.

\begin{definition}[Smoothness~\cite{D:JACM01}]
	A propositional formula $F$ is \textit{smooth} iff for each disjunction $D = D_1 \vee D_2 \vee \ldots \vee D_m$, the disjuncts have the same set of variables ($\forall i,j: vars(D_i) = vars(D_j)$).
\end{definition}

Smoothness is often enforced for disjunctions to simplify algorithms when using d-DNNFs~\cite{D:AAAI02,MMBH:AAAI10,DM:JAIR02,D:ANCL01}.
For a smooth disjunction, the sets of variables are equal for every disjunct by definition.
Without smoothness, if one of the disjuncts $D_i$ contains a variable $v$ that does not appear in another disjunct $D_j$, $v$ can be freely assigned (\ie, either $\top$ or $\bot$) in the context of $D_j$.
In this case, the number of satisfying assignments for $D_j$ needs to be multiplied by a factor of two for each missing variable.
Consequently, Line 5 in \autoref{algo:ddnnf:base} would require an adaptation to cope with free variables.
If we presume smoothness, the algorithm does not need to keep track of the set of variables.
According to Darwiche~\cite{D:JACM01}, every d\nobreakdash-DNNF can be smoothed in polynomial time with respect to its number of nodes and variables.
In this work, we consider a d-DNNF to be smooth if not stated otherwise.

Each literal induces exactly one satisfying assignment (\ie, setting the respective variable to $\top$ for $f$ or $\bot$ for $\neg f$) as seen in Lines 6--7 of \autoref{algo:ddnnf:base}.
Due to the presented properties, namely decomposability, determinism, and smoothness, the d-DNNF can be traversed with \autoref{algo:ddnnf:base} to compute the number of satisfying assignments of a propositional formula~\cite{D:ANCL01}.
The cardinality $\texttt{\#}FM$ of a feature model $\mathit{FM} = (\mathit{FT},CT)$ is equivalent to the number of overall satisfying assignments \texttt{\#}$\mathit{F_{FM}}$.
Hence, the algorithm can be directly used to compute the number of valid configurations for the entire feature model using a d-DNNF representing $\mathit{F_{FM}}$~\cite{SGRM:LPAIR18,SHN+:EMSE23}.
In \autoref{sec:concept}, we adapt the algorithm to compute the cardinalities of features and partial configurations.

\section{Reusing d-DNNFs for Cardinality of Features and Partial Configurations} \label{sec:concept}
In this work, we use the benefits of knowledge compilation to d-DNNF for feature-model counting.
In particular, we focus on repetitive queries as required for cardinality of features and partial configurations.
For a given feature model, we compile it to d-DNNF in a one-time offline phase using existing compilers~\cite{D:AAAI02,LM:IJCAI17,MMBH:AAAI10}.
Afterwards, we reuse the compiled d\nobreakdash-DNNF by reducing the two types of cardinalities to linear time (\wrt the number of d-DNNF nodes) queries to this d-DNNF.
The respective d\nobreakdash-DNNF queries shown in \autoref{sec:concept:queries} are inspired from ideas of an algorithm introduced by Darwiche~\cite{D:ANCL01} but have never been implemented nor evaluated.
Further, we propose several optimizations to those queries in \autoref{sec:concept:optimizations}.

\subsection{d-DNNF Queries for Features and Partial Configurations} \label{sec:concept:queries}

\begin{figure}[b]
	\tikzstyle{labelshift} = [xshift=0.75cm, yshift=-0.9cm]
	\tikzstyle{polygonlabelshift} = [xshift=0.85cm, yshift=-0.85cm]
	\tikzstyle{polygonorlabelshift} = [xshift=1.4cm, yshift=-0.7cm]

	\centering
	\begin{tikzpicture}[dag]
		\node [regular polygon,regular polygon sides=6,label={[polygonlabelshift] $4 \, \textcolor{colorcardfeature}{2} \, \textcolor{colorcardpc}{1}$}] {$\bm\land$}
		child[sibling distance= 20mm] {
				node [label={[labelshift] $1 \, \textcolor{colorcardfeature}{1} \, \textcolor{colorcardpc}{1}$}] {$A$}
			}
		child {
				node [orstyle,regular polygon,regular polygon sides=6,rotate=30,label={[polygonorlabelshift] $2 \, \textcolor{colorcardfeature}{1} \, \textcolor{colorcardpc}{1}$}] {\rotatebox{330}{$\bm\lor$}}
				child[sibling distance=25mm] {
						node [regular polygon,regular polygon sides=6,label={[polygonlabelshift] $1 \, \textcolor{colorcardfeature}{1} \, \textcolor{colorcardpc}{1}$}] {$\bm\land$}
						child {
								node [label={[labelshift] $1 \, \textcolor{colorcardfeature}{1} \, \textcolor{colorcardpc}{1}$}] {$B$}
							}
						child {
								node [inner sep = 1pt, label={[labelshift] $1 \, \textcolor{colorcardfeature}{1} \, \textcolor{colorcardpc}{1}$}] {$\lnot C$}
							}
					}
				child[sibling distance=35mm] {
						node [regular polygon,regular polygon sides=6,label={[polygonlabelshift] $1 \, \textcolor{colorcardfeature}{0} \, \textcolor{colorcardpc}{0}$}] {$\bm\land$}
						child {
								node [inner sep = 1pt, label={[labelshift] $1 \, \textcolor{colorcardfeature}{0} \, \textcolor{colorcardpc}{1}$}] {$\lnot B$}
							}
						child {
								node [inner sep = 1pt, label={[labelshift] $1 \, \textcolor{colorcardfeature}{1} \, \textcolor{colorcardpc}{0}$}] {$C$}
							}
					}
			}
		child[sibling distance= 42mm] {
				node [orstyle,regular polygon,regular polygon sides=6,rotate=30,label={[polygonorlabelshift] $2 \, \textcolor{colorcardfeature}{2} \, \textcolor{colorcardpc}{1}$}] {\rotatebox{330}{$\bm\lor$}}
				child {
						node [label={[labelshift] $1 \, \textcolor{colorcardfeature}{1} \, \textcolor{colorcardpc}{1}$}] {$D$}
					}
				child[sibling distance=17mm] {
						node [inner sep = 1pt, label={[labelshift] $1 \, \textcolor{colorcardfeature}{1} \, \textcolor{colorcardpc}{0}$}] {$\neg D$}
					}
			};
		\node[rectangle,align=left] at (4.3,0.2) {\small $\texttt{\#}FM$: $\cdot$ \small \phantom{dd} $\bm\wedge:$ $\prod$ \\ \small $\texttt{\#}f$: \textcolor{colorcardfeature}{$B$} \hspace{1.77em} \small $\bm\vee:$ $\sum$ \\ \small $\texttt{\#}C_p$: \textcolor{colorcardpc}{$\{\lnot C, D\}$}};
	\end{tikzpicture}
	\caption{Computing Cardinalities on a d-DNNF}
	\label{fig:algo:ddnnf}
\end{figure}
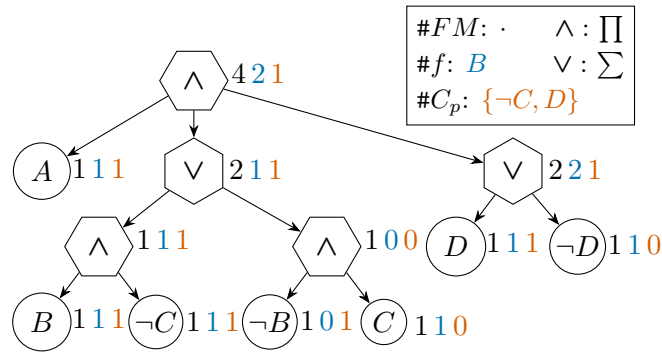

\autoref{algo:cardinality:feature} and \autoref{algo:cardinality:partialconfiguration} respectively depict the computation for cardinality of features and partial configurations via d-DNNF traversal.
Hereby, both algorithms adapt \autoref{algo:ddnnf:base} by performing different computations for the literals.
Note that Lines 1--6 are exactly equal for both algorithms.
\autoref{fig:algo:ddnnf} shows an example d-DNNF representing the formula $A \land ((B \land  \neg C) \lor (\neg B \land C)) \land (D \lor \neg D)$. Each node corresponds to a logical operator or literal.
The three numbers next to the nodes represent the number of satisfying assignments of that node for example queries of the three types of cardinalities. In \autoref{fig:algo:ddnnf}, the black numbers refer to traversing the d-DNNF to compute the overall number of satisfying assignments (equivalently: valid configurations) as seen in \autoref{algo:ddnnf:base}.
The result in root node of the d-DNNF in \autoref{fig:algo:ddnnf} indicates a cardinality of $\texttt{\#}\mathit{F_{FM}} = 4$.

\noindent
\begin{minipage}{0.49\textwidth}
	\raggedright
	\begin{algorithm}[H]
		\caption{Cardinality of Features \\~}
		\label{algo:cardinality:feature}
		\begin{algorithmic}[1]
			\Procedure{\#node}{node}
			\If {node is \texttt{And}}
			\State \textbf{return} $\prod_{c \in \texttt{children}}$ \Call{\#node}{c}
			\ElsIf{node is \texttt{Or}}
			\State \textbf{return} $\sum_{c \in \texttt{children}}$ \Call{\#node}{c}
			\ElsIf{node is \texttt{Literal}}
			\State \textbf{return} \Call{\#literal\_feat}{node}
			\EndIf
			\EndProcedure
			\Procedure{\#literal\_feat}{literal, feature}
			\If {literal.variable == feature}
			\If{literal.\texttt{positive}}{ \textbf{return} 1}
			\Else{ \textbf{return} 0}
			\EndIf
			\Else{ \textbf{return} 1}
			\EndIf
			\EndProcedure
			\State
			\State
			\State
			\State
		\end{algorithmic}
	\end{algorithm}
\end{minipage}
\hfill
\begin{minipage}{0.49\textwidth}
	\begin{algorithm}[H]
		\caption{Cardinality of Partial Configurations}
		\label{algo:cardinality:partialconfiguration}
		\begin{algorithmic}[1]
			\Procedure{\#node}{node}
			\If {node is \texttt{And}}
			\State \textbf{return} $\prod_{c \in \texttt{children}}$ \Call{\#node}{c}
			\ElsIf{node is \texttt{Or}}
			\State \textbf{return} $\sum_{c \in \texttt{children}}$ \Call{\#node}{c}
			\ElsIf{node is \texttt{Literal}}
			\State \textbf{return} \Call{\#literal\_pc}{node}
			\EndIf
			\EndProcedure

			\Procedure{\#literal\_pc}{literal, included, excluded}
			\If {literal.variable $\in$ included}
			\If{literal.\texttt{positive}}{ \textbf{return} 1}
			\Else{ \textbf{return} 0}
			\EndIf
			\ElsIf{literal.variable $\in$ excluded}
			\If{literal.\texttt{positive}}{ \textbf{return} 0}
			\Else{ \textbf{return} 1}
			\EndIf
			\Else
			\State \textbf{return} 1
			\EndIf
			\EndProcedure
		\end{algorithmic}
	\end{algorithm}
\end{minipage}

\paragraph{Features}
To compute the cardinality $\texttt{\#}f$ of a feature $f \in \mathit{FT}$ with $\mathit{FM} = (\mathit{FT},\mathit{CT})$, we need to exclude valid configurations $C = (I,E)$ with $C \in \mathit{VC}$ that do not include $f$ (\ie, $f \notin I$).
Therefore, when traversing the d-DNNF to compute $\texttt{\#}f$, we discard each satisfying assignment that falsifies $f$ by setting negative literals $\neg f$ to zero.
In \autoref{fig:algo:ddnnf}, the \textcolor{colorcardfeature}{blue} numbers refer the cardinality of $B$.
The literal $\neg B$ is initialized with a value of zero.
Positive literals of $B$ and other literals are initialized with a value of one.
The handling of literals for the cardinality of a feature is described in lines 8--12 of \autoref{algo:cardinality:feature}.
Using the adapted literal values, the number of satisfying assignments including $B$ can be computed by traversing the d-DNNF which results in a cardinality $\texttt{\#}f = 2$ as seen in the root of \autoref{fig:algo:ddnnf}.

\paragraph{Partial Configurations}
For the cardinality $\texttt{\#}C_p$ of a partial configuration $C_p = (I_p, E_p)$, we exclude assignments that assign false to an included feature $i \in I_p$ or true to an excluded feature $e \in E_p$.
In \autoref{fig:algo:ddnnf}, the \textcolor{colorcardpc}{orange} numbers indicate the cardinality for the partial configuration $\{\neg C, D\}$.
For each included feature (here: $D$), positive and negative literals are initialized with one and zero, respectively.
For each excluded feature (here: $C$), positive and negative literals are initialized with zero and one, respectively.
Literals that correspond to features that are neither included nor excluded are always initialized with a value of one.
The handling of literals for the cardinality of a partial configuration is specified in Lines 8--16 of \autoref{algo:cardinality:partialconfiguration}.
After initializing the values for literal nodes, the d-DNNF can be traversed to compute the cardinality $\texttt{\#}C_p = 1$ as seen in the root of \autoref{fig:algo:ddnnf}.

\paragraph{Correctness}
To show the correctness of our proposed \autoref{algo:cardinality:feature} and \autoref{algo:cardinality:partialconfiguration}, we presume that the original algorithm (\autoref{algo:ddnnf:base}) is correct~\cite{D:ANCL01,DM:JAIR02}.
The correctness of \autoref{algo:ddnnf:base} depends on the three properties decomposability, determinism, and smoothness.
When applying the assumptions for features and partial configurations, we essentially perform decision propagation for each variable of interest.
Alternatively stated, for a decision $d$ we replace each conflicting literal $\neg d$ with $\bot$ (false) and literals $d$ with $\top$ (true).
Hence, the adapted formula represents the subset of satisfying assignments that adheres to the assumptions.
For the correctness of \autoref{algo:cardinality:feature} and \autoref{algo:cardinality:partialconfiguration},
it remains to show that the adapted formula is still (1) decomposable, (2) deterministic, and (3) smooth after applying the assumptions.
First, decomposability is never violated as we only remove variables.
It follows that the sets of variables for each conjunct of an $\wedge$-node are still disjunct afterwards.
Second, each deterministic $\vee$-node is still deterministic.
With decision propagation, we never add new satisfying assignments but only remove them.
Hence, two set of satisfying assignments that were disjunct prior to applying the assumptions are still disjunct.
Third, for a smooth $\vee$-node, we remove the same set of variables for every child node.
Hence, if the set of variables were equal before applying the assumptions, the set of variables stay equal after.
Overall, traversing the adapted d-DNNF with \autoref{algo:cardinality:feature} and \autoref{algo:cardinality:partialconfiguration} yields correct results since the adapted formula represents the set of satisfying assignments conforming to the query and is still a smooth d-DNNF.

\paragraph{Summary}
With the presented algorithms, d-DNNFs can be reused to compute results for every application based on the cardinality of features and partial configurations.
Each described algorithm is an online computation on an already compiled d-DNNF and has a linear runtime with respect to the number of d-DNNF nodes.

\subsection{Optimizations} \label{sec:concept:optimizations}

For the algorithms presented in \autoref{sec:concept}, every node of a given d-DNNF needs to be traversed for each query.
Even though the complexity per query is linear in the number of nodes, analyses in practice may induce a high number of queries on d-DNNFs which may contain millions of nodes.
Further, the algorithm requires processing large numbers with up-to thousands of digits in each step~\cite{STS:VaMoS20}.
We propose several optimizations to (1) reduce the number of required traversals, (2) prevent evaluating equivalent nodes multiple times, and (3) decrease the share of d-DNNF nodes traversed per query.

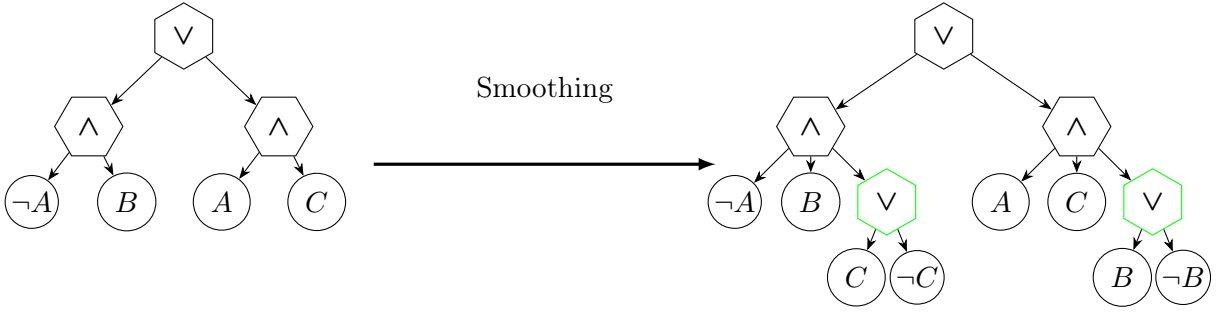
\begin{figure}
	\tikzstyle{labelshift} = [xshift=0.7cm, yshift=-0.85cm]
	\tikzstyle{polygonlabelshift} = [xshift=0.85cm, yshift=-0.85cm]
	\centering
	\begin{tikzpicture}[dag]
		\node [orstyle,regular polygon,regular polygon sides=6,rotate=30] (left) at (0,0) {\rotatebox{330}{$\bm\lor$}}
		child[sibling distance= 25mm] {
				node [regular polygon,regular polygon sides=6] {$\bm\land$}
				child {
						node [inner sep = 1pt] {$\neg A$}
					}
				child[sibling distance=10mm] {
						node {$B$}
					}
			}
		child[sibling distance= 25mm] {
				node [regular polygon,regular polygon sides=6] {$\bm\land$}
				child {
						node {$A$}
					}
				child[sibling distance=10mm] {
						node {$C$}
					}
			};

		\node [orstyle,regular polygon,regular polygon sides=6,rotate=30] (right) at (10,0) {\rotatebox{330}{$\bm\lor$}}
		child[sibling distance= 35mm] {
				node [regular polygon,regular polygon sides=6] {$\bm\land$}
				child[sibling distance=10mm] {
						node [inner sep = 1pt] {$\neg A$}
					}
				child[sibling distance=10mm] {
						node {$B$}
					}
				child[sibling distance=10mm] {
						node[orstyle,regular polygon,regular polygon sides=6,draw=green,rotate=30] {\rotatebox{330}{$\bm\lor$}}
						child[sibling distance = 8mm] {
								node {$C$}
							}
						child[sibling distance = 8mm] {
								node[inner sep = 1pt] {$\neg C$}
							}
					}
			}
		child[sibling distance= 35mm] {
				node [regular polygon,regular polygon sides=6] {$\bm\land$}
				child[sibling distance=10mm] {
						node {$A$}
					}
				child[sibling distance=10mm] {
						node {$C$}
					}
				child[sibling distance=10mm] {
						node[orstyle,regular polygon,regular polygon sides=6,draw=green,rotate=30] {\rotatebox{330}{$\bm\lor$}}
						child[sibling distance = 8mm] {
								node {$B$}
							}
						child[sibling distance = 8mm] {
								node[inner sep = 1pt] {$\neg B$}
							}
					}
			};

		\draw[-latex, very thick] (2.5,-1.7) -- node[above, draw=none, inner sep = 1pt, outer sep= 1pt] {Smoothing} (7,-1.7);

	\end{tikzpicture}
	\caption{Smoothing a d-DNNF}
	\label{fig:smoothing}
\end{figure}

\paragraph{Smoothing}
As discussed in \autoref{sec:background:ddnnf}, computing the cardinalities on an unsmooth d-DNNF causes computational overhead as the algorithm needs to keep track on the set of variables during the traversal~\cite{DM:JAIR02}.
To ensure smoothness, we add designated nodes for each unsmooth $\vee$-node as seen in \autoref{fig:smoothing} in a preprocessing step.
Hereby, for each missing variable $v \in \mathit{missingVariables}_\Phi$ in a child subtree $\Phi$ of the $\vee$-node, we add a disjunction $v \vee \neg v$ resulting in the following smooth disjunction on the right side.
$$\bigvee_{\Phi \in children} \Phi \qquad \rightarrow \qquad \bigvee_{\Phi \in children} (\Phi \wedge \bigwedge_{v \in \mathit{missingVariables_\Phi}} v \vee \neg v)$$

The added subformulas are tautologies as $v \lor \neg v \equiv \top$ holds for every variable $v$.
Hence, when adding smoothing nodes to a subformula $\Phi$, the result $\Phi \land \top$ equivalent to $\Phi$.
In the example \autoref{fig:smoothing}, the left subtree does not include the variable $C$.
Analogously, the right subtree does not include $B$.
Hence, the set of variables of the subtrees are not equal, which violates smoothness.
Thus, we conjunct a smoothing node $C \lor \neg C$ and $B \lor \neg B$ to the subtrees, respectively.
After the smoothing step, both sides of the root $\lor$ include the same set of variables $\{A, B, C\}$ without changing the semantics of the formula.

\begin{figure}
	\tikzstyle{labelshift} = [xshift=0.7cm, yshift=-0.85cm]
	\tikzstyle{polygonlabelshift} = [xshift=0.85cm, yshift=-0.85cm]
	\centering
	\begin{tikzpicture}[dag]
		\node [orstyle,regular polygon,regular polygon sides=6,rotate=30] (left) at (0,0) {\rotatebox{330}{$\bm\lor$}}
		child[sibling distance= 35mm] {
				node [regular polygon,regular polygon sides=6] {$\bm\land$}
				child[sibling distance=10mm] {
						node [inner sep = 1pt] {$\neg A$}
					}
				child[sibling distance=10mm] {
						node {$B$}
					}
				child[sibling distance=10mm] {
						node[orstyle, regular polygon,regular polygon sides=6,rotate=30] {\rotatebox{330}{$\bm\lor$}}
						child[sibling distance = 8mm] {
								node {$C$}
							}
						child[sibling distance = 8mm] {
								node[inner sep = 1pt] {$\neg C$}
							}
					}
			}
		child[sibling distance= 35mm] {
				node [regular polygon,regular polygon sides=6] {$\bm\land$}
				child[sibling distance=10mm] {
						node {$A$}
					}
				child[sibling distance=10mm] {
						node {$D$}
					}
				child[sibling distance=10mm] {
						node[orstyle, regular polygon,regular polygon sides=6,rotate=30] {\rotatebox{330}{$\bm\lor$}}
						child[sibling distance = 8mm] {
								node {$C$}
							}
						child[sibling distance = 8mm] {
								node[inner sep = 1pt] {$\neg C$}
							}
					}
			};

		\node [orstyle,regular polygon,regular polygon sides=6,rotate=30] (right) at (10,0) {\rotatebox{330}{$\bm\lor$}}
		child[sibling distance= 35mm] {
				node [regular polygon,regular polygon sides=6] {$\bm\land$}
				child[sibling distance=10mm] {
						node [inner sep = 1pt] {$\neg A$}
					}
				child[sibling distance=10mm] {
						node {$B$}
					}
				child[sibling distance=10mm] {
						node[orstyle,regular polygon,regular polygon sides=6,rotate=30] (reuse) {\rotatebox{330}{$\bm\lor$}}
						child[sibling distance = 8mm] {
								node {$C$}
							}
						child[sibling distance = 8mm] {
								node[inner sep = 1pt] {$\neg C$}
							}
					}
			}
		child[sibling distance= 35mm] {
				node[regular polygon,regular polygon sides=6] (reuseparent) {$\bm\land$}
				child[sibling distance=10mm] {
						node {$A$}
					}
				child[sibling distance=10mm] {
						node {$D$}
					}
			};

		\draw[-latex, very thick] (3.2,-1.7) -- node[above, draw=none, inner sep = 1pt, outer sep= 1pt] {Reuse Subtree} (7,-1.7);
		\draw[-latex,green] (reuseparent) -- (reuse);

	\end{tikzpicture}
	\caption{Reusing Subtrees of d-DNNF}
	\label{fig:optimization:reusing}
\end{figure}
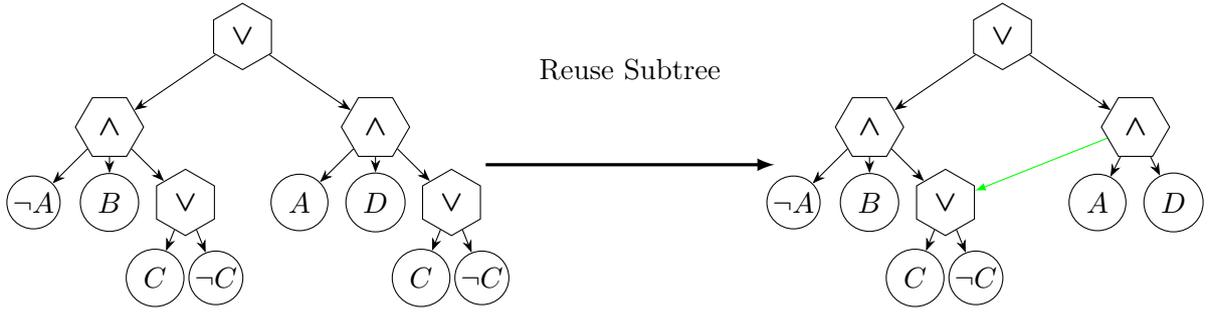

\paragraph{Reusing Subtrees}
In practice, equal subtrees are included several times in a d-DNNF~\cite{MMBH:AAAI10,LM:IJCAI17,D:AAAI02}.
Without reusing results for those subtrees during computation of our queries, each subtree instance would need to be evaluated separately.
\autoref{fig:optimization:reusing} shows a d-DNNF with an exact same subtree used two times on the left side.
When traversing the d-DNNF for computing the number of valid configurations, the corresponding subformula $C \lor \neg C$ needs to be evaluated twice.
To prevent computing the results for identical subtrees multiple times, we reuse subtrees as shown on the right side of \autoref{fig:optimization:reusing}.
In this case, the subtree corresponding to $C \lor \neg C$ has two parents.
After computing the number of satisfying assignments for a subtree with multiple parents (here: $C \lor \neg C$), the value is cached and reused for the remaining parent nodes.

% Note that each available d-DNNF compiler~\cite{D:AAAI02,LM:IJCAI17,MMBH:AAAI10} already ensures the described order when generating the d-DNNF.

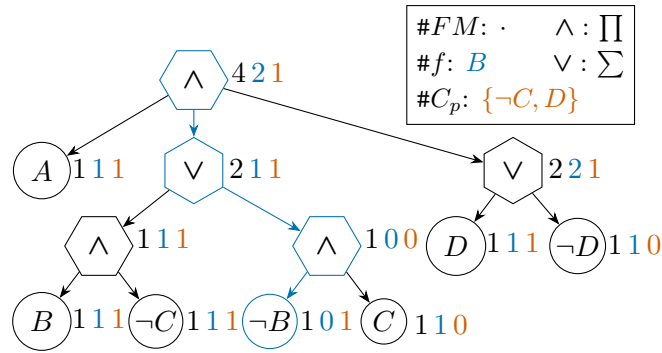
\begin{figure}[b]
	\tikzstyle{labelshift} = [xshift=0.75cm, yshift=-0.9cm]
	\tikzstyle{polygonlabelshift} = [xshift=0.85cm, yshift=-0.85cm]
	\tikzstyle{polygonorlabelshift} = [xshift=1.4cm, yshift=-0.7cm]
	\centering
	\begin{tikzpicture}[dag]
		\node [draw=colorcardfeature,regular polygon,regular polygon sides=6,label={[polygonlabelshift] $4 \, \textcolor{colorcardfeature}{2} \, \textcolor{colorcardpc}{1}$}] {$\bm\land$}
		child[sibling distance= 20mm] {
				node [label={[labelshift] $1 \, \textcolor{colorcardfeature}{1} \, \textcolor{colorcardpc}{1}$}] {$A$}
			}
		child[draw=colorcardfeature] {
				node [orstyle,regular polygon,regular polygon sides=6,rotate=30,label={[polygonorlabelshift] $2 \, \textcolor{colorcardfeature}{1} \, \textcolor{colorcardpc}{1}$}] {\rotatebox{330}{$\bm\lor$}}
				child[sibling distance=25mm,draw=black] {
						node [regular polygon,regular polygon sides=6,label={[polygonlabelshift] $1 \, \textcolor{colorcardfeature}{1} \, \textcolor{colorcardpc}{1}$}] {$\bm\land$}
						child {
								node [label={[labelshift] $1 \, \textcolor{colorcardfeature}{1} \, \textcolor{colorcardpc}{1}$}] {$B$}
							}
						child {
								node [inner sep = 1pt, label={[labelshift] $1 \, \textcolor{colorcardfeature}{1} \, \textcolor{colorcardpc}{1}$}] {$\lnot C$}
							}
					}
				child[sibling distance=35mm] {
						node [regular polygon,regular polygon sides=6,label={[polygonlabelshift] $1 \, \textcolor{colorcardfeature}{0} \, \textcolor{colorcardpc}{0}$}] {$\bm\land$}
						child {
								node [inner sep = 1pt, label={[labelshift] $1 \, \textcolor{colorcardfeature}{0} \, \textcolor{colorcardpc}{1}$}] {$\lnot B$}
							}
						child[draw=black] {
								node [inner sep = 1pt, label={[labelshift] $1 \, \textcolor{colorcardfeature}{1} \, \textcolor{colorcardpc}{0}$}] {$C$}
							}
					}
			}
		child[sibling distance= 42mm] {
				node [orstyle,regular polygon,regular polygon sides=6,rotate=30,label={[polygonorlabelshift] $2 \, \textcolor{colorcardfeature}{2} \, \textcolor{colorcardpc}{1}$}] {\rotatebox{330}{$\bm\lor$}}
				child {
						node [label={[labelshift] $1 \, \textcolor{colorcardfeature}{1} \, \textcolor{colorcardpc}{1}$}] {$D$}
					}
				child[sibling distance=17mm] {
						node [inner sep = 1pt, label={[labelshift] $1 \, \textcolor{colorcardfeature}{1} \, \textcolor{colorcardpc}{0}$}] {$\neg D$}
					}
			};
		\node[rectangle,align=left] at (4.3,0.2) {\small $\texttt{\#}FM$: $\cdot$ \small \phantom{dd} $\bm\wedge:$ $\prod$ \\ \small $\texttt{\#}f$: \textcolor{colorcardfeature}{$B$} \hspace{1.77em} \small $\bm\vee:$ $\sum$ \\ \small $\texttt{\#}C_p$: \textcolor{colorcardpc}{$\{\lnot C, D\}$}};
	\end{tikzpicture}
	\caption{Partial Traversal on a d-DNNF for the Cardinality of Feature $B$}
	\label{fig:optimization:partialtraversal}
\end{figure}

\paragraph{Partial Traversals}
If we compute the number of satisfying assignments under a partial assignment of literals (\eg, for the cardinality of a feature), we only have to recompute the value of nodes that are influenced by a corresponding literal. Hence, for each query, we first traverse the nodes of a d-DNNF to identify nodes that are part of a direct path from a corresponding literal to the root and mark these nodes. Afterwards, we traverse the d-DNNF and only recompute the count of marked nodes.
To reuse values that do not change due to a given partial assignment, we cache the original value (\ie, without partial assignment) for every node.
\autoref{fig:optimization:partialtraversal} shows the partial traversal (\textcolor{colorcardfeature}{blue} path) of computing the cardinality of $B$ for our running example.
The traversal only requires adapting the values from a direct path between $\neg B$ and the root $\land$.
Note that for every unaffected/uncolored node, the value for the cardinality of \textcolor{colorcardfeature}{$B$} and the original value is equal.
Hence, only four out of the twelve overall nodes need to re-evaluated with our partial traversal.

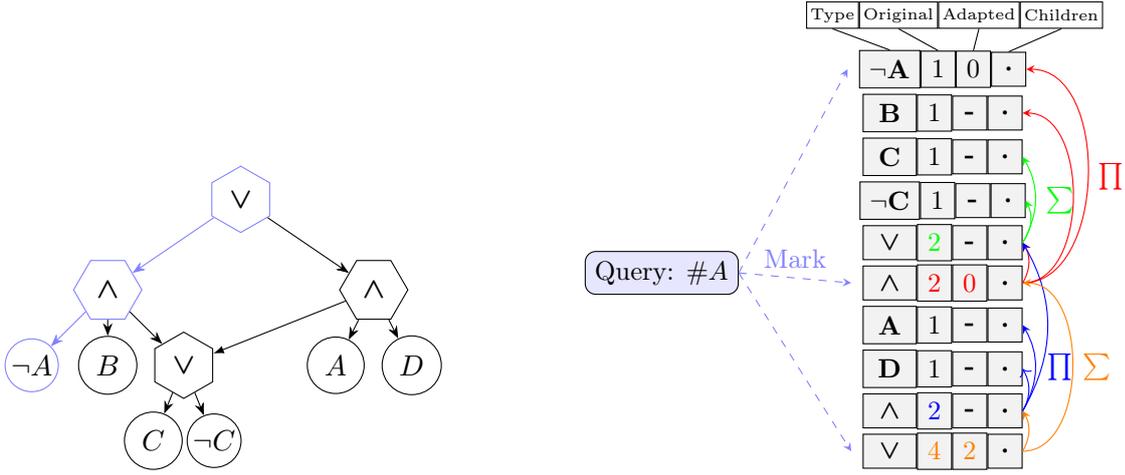
\begin{figure}
	\newcommand{\colorroot}{orange}
	\newcommand{\colorleftand}{red}
	\newcommand{\colorrightand}{blue}
	\newcommand{\colororc}{green}

	\tikzstyle{entrybox} = [fill=gray!10,draw=black,rectangle,minimum height= 0.45cm]
	\tikzstyle{nodebox} = [entrybox,minimum width=0.7cm]
	\tikzstyle{countbox} = [entrybox,minimum width=0.45cm]
	\tikzstyle{headerbox} = [draw=black,rectangle,minimum height= 0.35cm, inner sep = 1.5pt]

	\begin{tikzpicture}[dag]

		\node [orstyle,regular polygon,regular polygon sides=6,rotate=30, draw=blue!50] (right) at (-7,0) {\rotatebox{330}{$\bm\lor$}}
		child[sibling distance= 35mm,draw=blue!50] {
				node [regular polygon,regular polygon sides=6] {$\bm\land$}
				child[sibling distance=10mm] {
						node [inner sep = 1pt] {$\neg A$}
					}
				child[sibling distance=10mm,draw=black] {
						node {$B$}
					}
				child[sibling distance=10mm,draw=black] {
						node[orstyle, regular polygon,regular polygon sides=6,rotate=30] (reuse) {\rotatebox{330}{$\bm\lor$}}
						child[sibling distance = 8mm] {
								node {$C$}
							}
						child[sibling distance = 8mm] {
								node[inner sep = 1pt] {$\neg C$}
							}
					}
			}
		child[sibling distance= 35mm] {
				node[regular polygon,regular polygon sides=6] (reuseparent) {$\bm\land$}
				child[sibling distance=10mm] {
						node {$A$}
					}
				child[sibling distance=10mm] {
						node {$D$}
					}
			};

		\draw[-latex] (reuseparent) -- (reuse);
	\end{tikzpicture}%
	\qquad%
	\begin{tikzpicture}[font=\small]
		\node[nodebox] (litnegA) at (0,0) {$\mathbf{\neg A}$};
		\node[countbox, right = 0pt of litnegA] (countnegA) {1};
		\node[countbox,right = 0pt of countnegA] (tempnegA) {0};
		\node[countbox,right = 0pt of tempnegA] (pointerlitnegA) {$\bm\cdot$};

		\node[headerbox, above left= 8pt and 0pt of litnegA] (type) {\tiny Type};
		\node[headerbox, right = 0pt of type] (original) {\tiny Original};
		\node[headerbox, right = 0pt of original] (temp) {\tiny Adapted};
		\node[headerbox, right = 0pt of temp] (children) {\tiny Children};

		\draw (type.south) -- (litnegA.north);
		\draw (original.south) -- (countnegA.north);
		\draw (temp.south) -- (tempnegA.north);
		\draw (children.south) -- (pointerlitnegA.north);

		\node[nodebox,  below = 2pt of litnegA] (litB) {$\mathbf{B}$};
		\node[countbox, right = 0pt of litB] (countlitB) {1};
		\node[countbox,right = 0pt of countlitB] (templitB) {\textbf{-}};
		\node[countbox,right = 0pt of templitB] (pointerlitB) {$\bm\cdot$};

		\node[nodebox, below = 2pt of litB] (litC) {$\mathbf{C}$};
		\node[countbox,right = 0pt of litC] (countlitC) {1};
		\node[countbox,right = 0pt of countlitC] (templitC) {\textbf{-}};
		\node[countbox,right = 0pt of templitC] (pointerlitC) {$\bm\cdot$};

		\node[nodebox, below = 2pt of litC] (litnegC) {$\mathbf{\neg C}$};
		\node[countbox,right = 0pt of litnegC] (countlitnegC) {1};
		\node[countbox,right = 0pt of countlitnegC] (templitnegC) {\textbf{-}};
		\node[countbox,right = 0pt of templitnegC] (pointerlitnegC) {$\bm\cdot$};

		\node[nodebox, below = 2pt of litnegC] (orC) {$\bm\lor$};
		\node[countbox,right = 0pt of orC] (countorC) {\textcolor{\colororc}{2}};
		\node[countbox,right = 0pt of countorC] (temporC) {\textbf{-}};
		\node[countbox,right = 0pt of temporC] (pointerorC) {$\bm\cdot$};

		\node[nodebox, below = 2pt of orC] (leftand) {$\bm\land$};
		\node[countbox,right = 0pt of leftand] (countleftand) {\textcolor{\colorleftand}{2}};
		\node[countbox,right = 0pt of countleftand] (templeftand) {\textcolor{\colorleftand}{0}};
		\node[countbox,right = 0pt of templeftand] (pointerleftand) {$\bm\cdot$};

		\node[nodebox, below = 2pt of leftand] (litA) {$\mathbf{A}$};
		\node[countbox,right = 0pt of litA] (countlitA) {1};
		\node[countbox,right = 0pt of countlitA] (templitA) {\textbf{-}};
		\node[countbox,right = 0pt of templitA] (pointerlitA) {$\bm\cdot$};

		\node[nodebox, below = 2pt of litA] (litD) {$\mathbf{D}$};
		\node[countbox,right = 0pt of litD] (countlitD) {1};
		\node[countbox,right = 0pt of countlitD] (templitD) {\textbf{-}};
		\node[countbox,right = 0pt of templitD] (pointerlitD) {$\bm\cdot$};

		\node[nodebox, below = 2pt of litD] (rightand) {$\bm\land$};
		\node[countbox,right = 0pt of rightand] (countrightand) {\textcolor{\colorrightand}{2}};
		\node[countbox,right = 0pt of countrightand] (temprightand) {\textbf{-}};
		\node[countbox,right = 0pt of temprightand] (pointerrightand) {$\bm\cdot$};

		\node[nodebox, below = 2pt of rightand] (root) {$\bm\lor$};
		\node[countbox,right = 0pt of root] (countroot) {\textcolor{\colorroot}{4}};
		\node[countbox,right = 0pt of countroot] (temproot) {\textcolor{\colorroot}{2}};
		\node[countbox,right = 0pt of temproot] (pointerroot) {$\bm\cdot$};

		\draw[-stealth,\colororc] (pointerorC.east) to [bend right] node[right,midway] {$\sum$} (pointerlitC.east);
		\draw[-stealth,\colororc] (pointerorC.east) to [bend right] (pointerlitnegC.east);

		\draw[-stealth,\colorleftand] (pointerleftand.east) to [bend right]  (pointerorC.east);
		\draw[-stealth,\colorleftand] (pointerleftand.east) to [in=0, out=0] node[right,midway] {$\prod$} (pointerlitnegA.east);
		\draw[-stealth,\colorleftand] (pointerleftand.east) to [in=0, out=0] (pointerlitB.east);

		\draw[-stealth,\colorrightand] (pointerrightand.east) to [bend right]  (pointerlitA.east);
		\draw[-stealth,\colorrightand] (pointerrightand.east) to [bend right] node[right,midway, xshift=-4pt, yshift=-15pt] {$\prod$} (pointerorC.east);
		\draw[->,\colorrightand] (pointerrightand.east) to [bend right] (pointerlitD.east);

		\draw[-stealth,\colorroot] (pointerroot.east) to [bend right]  (pointerrightand.east);
		\draw[-stealth,\colorroot] (pointerroot.east) to [in=0, out=0] node[right,midway] {$\sum$} (pointerleftand.east);

		\node[rounded corners, fill=blue!10, draw=black] (query) at (-3,-2.7) {Query: $\#A$};

		\draw[-stealth,dashed,blue!50] (query.east) -- ([shift={(-0.15,0)}]litnegA.west);
		\draw[-stealth,dashed,blue!50] (query.east) --  node[midway, above] {Mark} ([shift={(-0.15,0)}]leftand.west);
		\draw[-stealth,dashed,blue!50] (query.east) -- ([shift={(-0.15,0)}]root.west);

		\node[draw=none] at (-5,0) {};

	\end{tikzpicture}
	\caption{List Format for Iterative Traversal Representing \autoref{fig:optimization:reusing}}
	\label{figure:optimizations:iterativetraversal}
\end{figure}

\paragraph{Iterative Traversal}
To simplify reusing values and overall reduce the computational effort when traversing the d-DNNF, we use an iterative approach instead of a recursive one.
The nodes are stored in a list such that a child is always traversed before its parents.
Thus, when processing a node the values of its children are already available and do not require any recursive descent.
\autoref{figure:optimizations:iterativetraversal} shows the list representation we use for the respective d-DNNF seen on the left side.
From left to right, the list representation shows the type of the node (\ie, $\wedge$), the original count, the adapted count, and pointers to its children.
To evaluate a query (here: $\#A$) on the d-DNNF, the list is traversed iteratively starting from the first entry in the list.
In this case, we start with the node representing the literal $\neg A$.
When reaching one of the nodes depending on the values of other nodes, all required values are already computed.
For instance, when reaching the first \textsc{or} node, we just look up the values for $C$ and $\neg C$.

For partial traversals, we ensure only visit marked nodes.
The dashed blue lines indicate which nodes were marked during the first phase for the query $\#A$.
Afterwards, the algorithm recomputes the value for each of the three marked nodes.
The result for the query $\#A = 2$ is then saved in the temp value of the root node.

\paragraph{Partial Calculations}
When computing the value of a node during a partial traversal, it is possible only for a small subset of its child nodes to be changed.
In this case, it may be beneficial to not recompute the entire node but adapt the cached value.
For instance, given a $\wedge$-node with ten children for which only one child $c'$ changed, we do not build the product of the ten children.
Instead, we adapt the original value based on the changed values as seen in \autoref{equation:folding:and}.
If more than half of the children changed, we recompute the value as shown in \autoref{algo:ddnnf:base}.
Otherwise, we divide the cached value by the old value of $c_{\mathit{old}}$ and then multiply it with the new value for every changed child.
Analogously, we subtract the cached value by the old value of $c_{\mathit{old}}$ and then add the new value for $\lor$ nodes as seen in \autoref{equation:folding:or}.
The partial calculation require two arithmetic operations per changed node.
For instance, recomputing an $\land$-node requires one multiplication and one division.
Hence, we only apply partial calculation if less than half of the children values changed.

\begin{equation}
	\#node(\land) = \begin{cases}
		\prod_{c \in \mathit{children}} \#c_{\mathit{new}}                                                                                      & \geq \frac{n}{2} \text{ children changed} \\
		\frac{\#\land_{\mathit{old}} \cdot \prod_{c \in \mathit{adapted}} \#c_{\mathit{new}}}{\prod_{c \in \mathit{adapted}}\#c_{\mathit{old}}} & \text{else}                               \\
	\end{cases}
	\label{equation:folding:and}
\end{equation}

\begin{equation}
	\#node(\lor) = \begin{cases}
		\sum_{c \in \mathit{children}} \#c_{\mathit{new}}                                             & \geq \frac{n}{2} \text{ children changed} \\
		\#\lor_{\mathit{old}} + \sum_{c \in \mathit{adapted}} \#c_{\mathit{new}} - \#c_{\mathit{old}} & \text{else}                               \\
	\end{cases}
	\label{equation:folding:or}
\end{equation}

\paragraph{Core and Dead Features}
Core and dead features appear in all and no valid configurations, respectively~\cite{BSRC10}.
Hence, the cardinality of core and dead features can be directly derived without traversing the d-DNNF again as seen in \autoref{equation:coredead}.
For a core feature, the number of satisfying assignments is always equal to the number of overall satisfying assignments.
The cardinality of a dead feature is always zero. 
As optimization for \ddknnife{}, we propose to pre-compute core and dead features by exploiting the smoothness of a d-DNNF to potentially reduce the number of required traversals.
With this novel proposal, we can identify all core and dead features with \textit{one} traversal through the literal nodes using the following properties.
A feature $f$ is core iff only positive literals $f$ appear in the d-DNNF.
Analogously, a feature $f$ is dead iff only negative literals $\neg f$ appear.
In our running example \autoref{fig:algo:ddnnf}, the feature $A$ is core as only positive literals appear in the smooth d-DNNF.

\begin{equation}
	\#f = \begin{cases}
		\#\mathit{FM} & \text{if } f \text{ is core} \\
		0             & \text{if } f \text{ is dead} \\
		\#f           & else
	\end{cases}
	\label{equation:coredead}
\end{equation}

In the following, we provide a short intuition on the correctness of our approach of detecting core and dead features.
Assuming that our algorithms for the cardinality of partial configurations are correct, the correctness of our proposal using core/dead features is directly implied as explained in the following.
We start the explanations with core features.
If only positive literals of a variable $v$ appear, applying \autoref{algo:cardinality:feature} for $\#v$ (\ie, cardinality of $v$) always produces the same results as the d-DNNF under no assumptions.
The only case resulting in a change would be a negative literal of $v$ which conflicts our assumptions.
Hence, the number of satisfying assignments $v$ appears in the is equal to the number of overall satisfying assignments making $v$ a core feature.
This works analogously for dead features.
If only negative literals of a variable $w$ appear, evaluating $\#(\neg w)$ with \autoref{algo:cardinality:feature} always yields the same result as for the d-DNNF without assumptions.
Thus, there are no satisfying assignments including $w$ in this case as otherwise the results would differ.

The knowledge about core and dead features can also be used to reduce the number of queries required for partial configurations.
Each time a dead feature is included or a core feature is excluded in a partial configuration, this configuration is unsatisfiable and does not need to be evaluated with a d-DNNF traversal.
Further, included core features and excluded dead features can be disregarded for the marking phase of partial traversals.

\paragraph{Summary}
Each of the six presented optimizations reduces either the number of required queries or reduces effort for a single traversal.
Both of those impacts potentially improves the efficiency of reusing d-DNNFs for computing the cardinality of features or partial configurations.
In \autoref{sec:evaluation}, we examine the general performance of our approach and also take a closer look on the effect of specific optimizations.

\section{Tool Support: ddnnife}\label{sec:toolsupport}

The presented algorithms and optimizations are implemented in our open-source d-DNNF reasoner \ddknnife{}, the Swiss knife for reusing d-DNNFs. \ddknnife{} is implemented in Rust\footnote{\url{https://www.rust-lang.org/}} as Rust generally compiles to efficient target code and has great support for improving memory safety. The source code and example input data is available at GitHub\footnote{\url{https://github.com/SoftVarE-Group/d-dnnf-reasoner/releases/tag/March2023}} under LPGLv3\footnote{\url{https://www.gnu.org/licenses/lgpl-3.0.de.html}} license allowing potential users to employ \ddknnife{} with only few restrictions even in commercial settings. \ddknnife{} takes a d-DNNF as input following available standard formats (\cf \autoref{sec:toolsupport:parsing}).
When parsing the d-DNNF, \ddknnife{} can be used to smooth the d-DNNF (if necessary) and compute the cardinality of the feature model, features, or partial configurations.
We use \ddknnife{}, amongst other tools, to evaluate the advantages of reusing d-DNNFs in \autoref{sec:evaluation}.

\subsection{Parsing d-DNNFs} \label{sec:toolsupport:parsing}

\paragraph{Input Format} \ddknnife{} supports the standard format set by \ctod{}\footnote{\url{http://reasoning.cs.ucla.edu/c2d/}}~\cite{D:AAAI02} and the adaptation of the format introduced by \dfour{}~\cite{LM:IJCAI17}.\footnote{\url{https://github.com/crillab/d4}} Supporting both formats, \ddknnife{} is able to parse d-DNNFs from each popular d-DNNF compiler, namely \ctod{}~\cite{D:AAAI02}, \dsharp{}~\cite{MMBH:AAAI10}, and \dfour{}~\cite{LM:IJCAI17}.

\autoref{listing:c2d} shows our running example \autoref{fig:algo:ddnnf} in the \ctod{} format.\footnote{\url{http://reasoning.cs.ucla.edu/c2d/}} The first line \texttt{nnf v e n} indicates the number of nodes (\texttt{v}), the number of edges (\texttt{e}), and the number of variables (\texttt{n}).
Each remaining line describes one node in the d-DNNF. \texttt{L 1} introduces a positive literal of the variable \texttt{1}. An and node is described by the line \texttt{A n i j k} with \texttt{n} children and \texttt{i}, \texttt{j}, \texttt{k} being indices of the lines holding child nodes.
Note that the first line containing a node has the index zero.
Or nodes come with an additional parameter that indicates a decision variable (\ie, a variable that is always true in the left and false in the right child). \texttt{O d n i j k} describes an or node with the decision variable \texttt{d}, \texttt{n} child nodes with the indices \texttt{i}, \texttt{j}, and \texttt{k}. A decision variable \texttt{d} of $0$ indicates that there is no decision variable.

\autoref{listing:d4} shows our running example \autoref{fig:algo:ddnnf} in the \dfour{} format.\footnote{\url{https://github.com/crillab/d4}}
Here, each line either describes a node or an edge between nodes and is terminated with a \texttt{0}.
Lines describing nodes always first specify the type of node, namely or (o), and (a), true (t), false (f), and then the index of this node.
In constract to the \ctod{}-format, the indices do not refer to the line index but to the index specified for each node.
Edges always follow the format \texttt{p c x y z 0} with the parent node \texttt{p}, the child node \texttt{c}, and an arbitrary number of literals \texttt{x y z} set to true when following this path.
For instance, Line 4 specifies that the node with the index \texttt{1} has a child ($\Phi_3$) with the index \texttt{3}.
Further, the literals \texttt{2} and \texttt{-3} are conjoined resulting in the subformula $(\Phi_3 \land (2 \land -3)) \lor \Phi_{\mathit{rest}}$ with $\Phi_{\mathit{rest}}$ being the formula resulting from other edges starting from node \texttt{1}.

\begin{minipage}[t]{0.45\textwidth}
	\begin{lstlisting}[numbers=none,language=ctodformat,caption={c2d Format},label={listing:c2d},basewidth = {.48em}]
     nnf 11 11 4
 0:  L 1 # A
 1:  L 2 # B
 2:  L -3 # !C
 3:  L -2 # !B
 4:  L 3 # C
 5:  L 4 # D
 6:  L -4 # !D
 7:  A 2 1 2
 8:  A 2 3 4
 9:  O 0 2 7 8
10:  O 4 2 5 6
11:  A 3 0 9 10
	\end{lstlisting}

\end{minipage}\hfill
\begin{minipage}[t]{0.45\textwidth}
	\begin{lstlisting}[language=dfourformat,caption={d4 Format},label={listing:d4}]
o 1 0
o 2 0
t 3 0
a 4 0
1 3 2 -3 0
1 3 -2 3 0
2 3 4 0
2 3 -4 0
4 3 1 0
4 0 2 0
4 0 1 0
	\end{lstlisting}
\end{minipage}

\begin{figure}[htbp]
	\centering
	\includegraphics[width=0.7\textwidth]{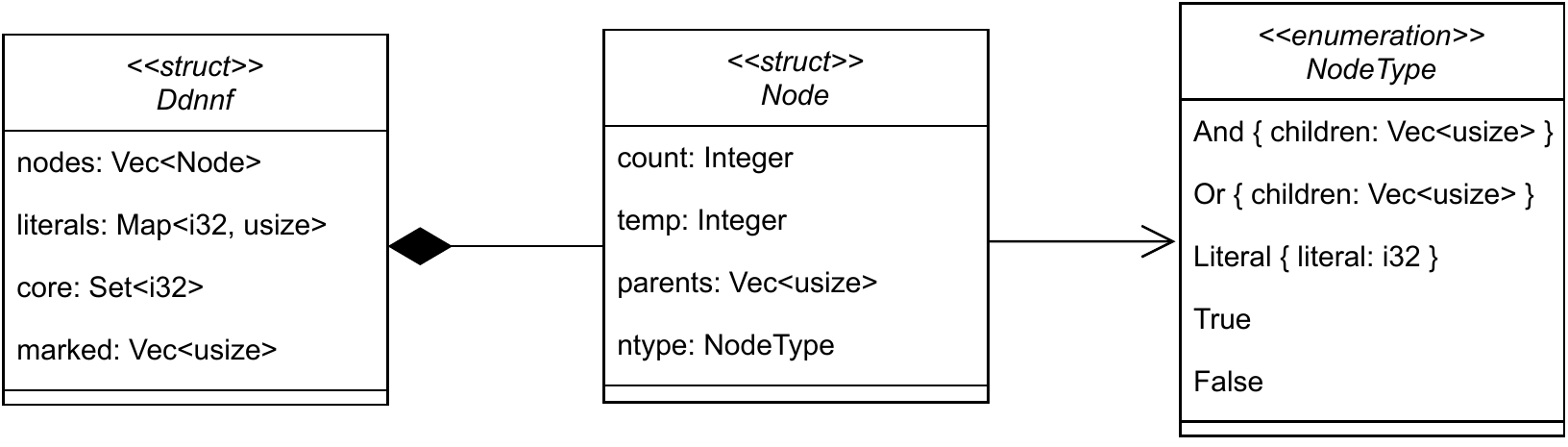}
	\caption{\ddknnife{} Data Structure}
	\label{figure:tool:datastructure}
\end{figure}

\paragraph{Data Structure} \autoref{figure:tool:datastructure} shows the data structure used in \ddknnife{} to store a d-DNNF.
Each d-DNNF mainly consists of a set of nodes with some additional information to simplify algorithms.
The additional information consist of (1) pointers to the indices of literal nodes, (2) the set of core literals, and (3) a list indicating nodes that need to be re-computed during a partial traversal.
The nodes have \texttt{NodeType}-enumeration indicating the type of d-DNNF node, namely \texttt{And}, \texttt{Or}, \texttt{Literal}, \texttt{True}, and \texttt{False}.
Each node contains a pointer to all parent nodes to simplify traversing from the bottom when marking for partial traversals.
To reuse values, we attached the original count (\ie, without assumptions) and the \texttt{temp} value under the current query.
\textsc{And} and \textsc{Or} nodes also provide a set of indices pointing to their child nodes.

\paragraph{Pre-Processing}
Prior to performing queries for features or partial configurations, \ddknnife{} performs some pre-processing steps to simplify the computation for those queries.
First, the given d-DNNF is smoothed by adding smoothing nodes if necessary (\cf \autoref{sec:concept:optimizations}).
Second, connections from child to parent node are added to simplify the marking algorithm for partial traversals.
Third, pointers to the indices of literal nodes are saved to be used as starting point for marking.
Fourth, core and dead features are identified by examining if a variable only appears as positive (core) or negative (dead) literal, respectively.
Fifth, the model count for each d-DNNF node is computed under no assumptions.
Those values are then re-used for partial traversals.
After the different the pre-processing steps, the resulting representation of the d-DNNF is then stored and used for querying.

\subsection{Implementation of Algorithms}

\ddknnife{} supports the algorithms for computing the cardinality of feature models, features, and partial configurations presented in \autoref{sec:concept:queries}. As the parsing step ensures that each d-DNNF is smooth, the algorithms do not keep track of variables during the computation.
The implementation of \autoref{algo:ddnnf:base} is slightly adapted to incorporate $\top$ and $\bot$ nodes as they are used by 
\dsharp{}~\cite{LM:IJCAI17} and \dfour{}~\cite{MMBH:AAAI10} in some cases.
$\top$ and $\bot$ are handled similarly literals with $\top$ always having a value of one and $\bot$ of zero.

\paragraph{Optimizations}
\ddknnife{} includes all optimzations presented in \autoref{sec:concept:optimizations}.
However, some optimizations are slightly adapted according to observations when applying \ddknnife{} to d-DNNFs representing feature models.

Folding is only applied to $\land$-nodes, due to the following two reasons.
First, multiplications are more computationally expensive than addition resulting in smaller runtime profits when reducing the number of additions.
Second, in practice, $\lor$-nodes typically have exact two child nodes due to construction of d-DNNFs with \ctod{}, \dsharp{}, and \dfour{}.
For exactly two children, the folding optimization provides no benefit as still two arithmetic operations are required when only one child changes.

When dealing with large partial configurations (\ie, many included or excluded features), a large portion of the d-DNNF changes and needs to be traversed.
Thus, if the size of the partial configurations exceeds a certain threshold, the overhead of marking nodes to recompute is no longer worth it.
In this case, we do not use partial traversals but re-evaluate the whole d-DNNF.

\subsection{Usage}
Prior to employing different analyses, \ddknnife{} always demands a d-DNNF, either in \ctod{} or \dfour{} format, as input.
In this section, we shortly explain the different modes of usage for \ddknnife{}, namely a one-time computation for a given set of queries and streaming mode for interactive usage.
For more details on syntax and technical features, we refer to the repository.\footnote{\url{https://github.com/SoftVarE-Group/d-dnnf-reasoner/releases/tag/March2023}}

\paragraph{One-Time Computation}
When invoking d-DNNF as one-time computation, the user has three options to denote queries: (1) compute cardinality of a feature, (2) cardinality of a partial configuration, (3) cardinality of all features, or (3) a set of different queries given in a text file.
After invocation, the tool provides the results either as \texttt{.csv} or command line output and then terminates.

% \begin{lstlisting}[language=bash,frame=single,caption={\ddknnife{} Example Invocation},label={listing:usage:onetime}]
% 	./ddnnife model.nnf --card_of_fs
% \end{lstlisting}

\paragraph{Streaming Mode}
The streaming mode of \ddknnife{} can be used for interactive querying if not all queries are known in advance.
At the start \ddknnife{} loads and pre-processes the d-DNNF in a one-time effort and waits for queries on features or partial configurations.
The streaming mode follows a strict protocol to allow users integrating \ddknnife{} in non-rust based tools via protocol-based exchange.
Nevertheless, \ddknnife{} can also be used interactively with human input.
After loading, the user can query the d-DNNF (\eg, \texttt{count v 1} to count the cardinality of variable 1).
For details on the protocol, we refer to the repository.\footnote{\url{https://github.com/SoftVarE-Group/d-dnnf-reasoner/releases/tag/March2023}}
% \begin{lstlisting}[language=bash,frame=single,caption={\ddknnife{} Example Interactive Usage},label={listing:usage:interactive}]
% 	./ddnnife model.nnf --stream 
% \end{lstlisting}

\paragraph{Options}
Besides usage mode and parameters for specifying queries, \ddknnife{} provides some additional parameters to control its behavior.
Often, pre-processing and in particular smoothing the d-DNNF requires a considerable initial effort.
Hence, we provide an option to save the smoothed d-DNNF for later usage.
Multithreading can also be enabled by the user to answer multiple queries in parallel.
In some cases, \dfour{} omits fully optional features in the d-DNNF.
When disregarding those variables, two problems emerge.
First, the number of satisfying assignments would be off by two to the power of omitted variables as those can be assigned arbitrarily.
Second, \ddknnife{} could not compute the cardinality of those features or partial configuration including/excluding one of those as they are not part of the d-DNNF.
Hence, we provide a parameter for declaring omitted variables to prevent faulty results.

\paragraph{Summary}
For a given d-DNNF, \ddknnife{} can be used interactively or with a one-time computation to perform repetitive counting queries.
With popular compilers, such as \dfour{}~\cite{LM:IJCAI17}, \ctod{}~\cite{D:AAAI02}, and \dsharp{}~\cite{MMBH:AAAI10}, we can translate a feature model to d-DNNF to perform feature-model counting with \ddknnife{}.
In \autoref{sec:evaluation}, we examine the performance of \ddknnife{} and its optimizations when applied to feature models.
Hereby, we focus on the mode for one-time computations.

\section{Evaluation}\label{sec:evaluation}
In our empirical evaluation, we analyze the benefits of reusing d-DNNFs for computing the cardinality of features and partial configurations compared to the state of the art.
The applicability of applying d-DNNFs for feature-model counting depends on the scalability of both the compilation (offline phase) and the reuse of the compiled d-DNNF with a reasoner (online phase).
Hence, we consider both phases in our empirical evaluation.
We provide a publicly available replication package\footnote{\url{https://github.com/SoftVarE-Group/exploiting-ddnnfs-eval/releases/tag/March2023}} that includes all evaluated tools, subject systems (except nine confidential systems of our industry partner), and the required tooling for measurements.

% In \autoref{subsec:eval:researchquestions}, we present and discuss our research questions.
% In \autoref{subsec:eval:experimentdesign}, we explain the experiment design used to gather insights to answer the research questions.
% Hereby, we describe the subject systems, the evaluated tools, the performed experiments, applied statistical tests, and the technical setup.
% In \autoref{subsec:eval:results}, we present the results of our experiments.
% In \autoref{subsec:eval:discussion}, we use the results to elaborate on our research questions.
% In \autoref{subsec:eval:threats}, we discuss potential threats to the validity of our results.

\subsection{Research Questions} \label{subsec:eval:researchquestions}
With our research questions, we aim to inspect the advantages of reusing d-DNNFs compared to repetitive \ssat{} calls.
Additionally, we compare the performance of publicly available tools.

\newcommand{\rqone}{How efficient are d-DNNF compilers when applied to industrial feature models?}
\newcommand{\rqtwo}{Which tools and tool combinations compute correct cardinalities?}
\newcommand{\rqthree}{How efficient is reusing d-DNNFs for computing cardinalities of features and partial configurations compared to repetitive \ssat{} calls?}
\newcommand{\rqfour}{How much do the optimizations in \ddknnife{} improve the efficiency regarding runtime?}

\begin{description}[leftmargin=\researchquestionlabelwidthof{1}]
	\researchquestion{1} \rqone{}
\end{description}

To apply our algorithms on industrial feature models, it is necessary to be able to compile the respective formulas to d-DNNF first.
With RQ1, we inspect and compare the scalability of d-DNNF compilers regarding runtime and size of the resulting d-DNNFs.
We also consider the size as the runtime of \ddknnife{} depends on the size of the d-DNNF.

\begin{description}[leftmargin=\researchquestionlabelwidthof{2}]
	\researchquestion{2} \rqtwo{}
\end{description}

To examine the correctness of considered tools, we compare computed cardinalities.
In particular, we empirically validate the correctness of our proposal \ddknnife{}.

\begin{description}[leftmargin=\researchquestionlabelwidthof{3}]
	\researchquestion{3} \rqthree{}
\end{description}

After examining the general applicability of applying d-DNNFs for feature-model counting with RQ1 and RQ2, we inspect the main focus of this work (\ie, advantages of reusing d-DNNFs for feature-model counting) with RQ3.
We compare runtimes of d-DNNF-based approaches with the runtimes of repetitive \ssat{} calls for computing cardinality of features and partial configurations.
Hereby, we consider reusing d-DNNFs with and without compilation time for comparison.
Further, we inspect after how many computations the d-DNNF exploitation hits a break-even point compared to repetitive \ssat{} calls.

% \begin{description}[leftmargin=\researchquestionlabelwidthof{4}]
% 	\researchquestion{4} \rqfour{}
% \end{description}

% Depending on task and feature model the tools may perform differently well regarding the runtime.
% For RQ4, we inspect the runtimes of available tools and tool combinations, including our proposal \ddknnife{}, for computing the cardinality of features and partial configurations for industrial feature models. \todo{Maybe cut}

\begin{description}[leftmargin=\researchquestionlabelwidthof{5}]
	\researchquestion{4} \rqfour{}
\end{description}

While the proposed base algorithms require only linear time \wrt to the number of d-DNNF nodes per query, the base algorithm may still require to perform arithmetic operations on millions of nodes for large d-DNNFs.
Our optimizations (\cf \autoref{sec:concept:optimizations}) aim to reduce the effort per query and the number of required queries. For RQ4, we inspect the effect of the proposed optimizations on the runtime of \ddknnife{}.

\subsection{Experiment Design} \label{subsec:eval:experimentdesign}
\paragraph{Subject Systems}
In our empirical evaluation, we only consider industrial feature models and no artificial ones.
Artificially generated feature models may result in runtimes that are not representative for the real world as observed previously~\cite{ABL:CP09,HST:SPLC21}.
With our selection of subject systems, we aim for a wide coverage of industrial feature models from a variety of domains.
To this end, we mostly use publicly available feature models used in benchmarks in related work.
\autoref{table:subjectsystems} gives an overview on the subject systems considering domain, number of features, number of clauses, and cardinality.

First, we include feature models from a benchmark\footnote{\url{https://github.com/AlexanderKnueppel/is-there-a-mismatch}} provided by Knüppel~\etal~\cite{KTS+:ESECFSE17} which consists of various CDL\footnote{\url{https://ecos.sourceware.org/}} models, KConfig\footnote{\url{https://www.kernel.org/doc/html/latest/kbuild/kconfig-language.html}} models, and an automotive product line.
For the 117 CDL feature models, we observed in previous work~\cite{SHN+:EMSE23} that they are highly similar considering various metrics, including number of features, number of constraints, and also performance of \ssat{} solvers.
To prevent a bias, we decided to include only three CDL feature models instead of all 117 in our benchmark.
Here, we took the feature models  with the minimum, median, and maximum number of features.
Second, our benchmark includes feature models from the systems software and application domain\footnote{\url{https://github.com/jeho-oh/Smarch}} provided by Oh~\etal~\cite{OGB+:TR20}.
Third, we include a version of the BusyBox feature model\footnote{\url{https://github.com/PettTo/Measuring-Stability-of-Configuration-Sampling}} provided by Pett~\etal~\cite{PKT+:SPLC21}.
Fourth, we consider a feature model\footnote{\url{https://github.com/PettTo/SPLC2019_The-Scalability-Challenge_Product-Lines}} extracted from the financial services domain~\cite{NMS+:GPCE18,PTR+:SPLC19}.
Fifth, we consider another feature model from the automotive domain published by Kowal~\etal~\cite{KAT:GPCE16}.
Sixth, we include a database feature model publicly available in the FeatureIDE examples\footnote{\url{https://github.com/FeatureIDE/FeatureIDE/tree/develop/plugins/de.ovgu.featureide.examples/featureide_examples}}.
Sixth, we include nine feature models representing automotive product lines from an industry collaboration, which onfortunately cannot be published.
Overall, our benchmark consists of 62 industrial feature models.

\begin{table}[]
	\caption{Overview Subject Systems}
	\label{table:subjectsystems}
	\setlength\tabcolsep{19.5pt}
	\begin{tabular*}{\textwidth}{@{}lrrrr}
		\toprule
		Domain             & Models & Features    & Clauses        & Cardinality                          \\ \midrule
		Operating System   & 37     & 94--62,482  & 190--343,944   & $10^{10}$--$10^{417}$\tiny$^\dagger$ \\
		Other Software$^*$ & 13     & 11--31,012  & 1--102,705     & $10^{3}$--$10^{5}$\tiny$^\dagger$    \\
		Automotive         & 11     & 384--18,616 & 1,020--350,221 & $10^{4}$--$10^{1534}$\tiny$^\dagger$ \\
		Financial Services & 1      & 771         & 7,241          & $10^{13}$                            \\ \bottomrule
	\end{tabular*}
	\newline\footnotesize $^*$ Including archivers (2),  database (2), video(2), network (1),  garbage collection (1), image (1), compiler (1), solving (1), web development (1), and team management (1)
	\newline\raggedright $^\dagger$ Cardinalities are incomplete due to solvers hitting timeouts for some models
\end{table}

\paragraph{Evaluated Tools}
For the selection of tools, our goal is to compare d-DNNF compilation and reuse to the state-of-the-art of feature-model counting.
Currently, the most scalable solution that is generally applicable appears to be translating a feature model to CNF and invoke a \ssat{} solver as black box~\cite{STS:VaMoS20,OGB+:TR19,MOP+:SPLC19}.
For representative results, we aim to include each publicly available d-DNNF compiler~\cite{D:AAAI02,LM:IJCAI17,MMBH:AAAI10} (\ie, tools compiling CNF to d-DNNF) and reasoner (\ie, tools using d-DNNF to compute number of satisfying assignments) and the fastest publicly available off-the-shelf \ssat{} solvers~\cite{STS:VaMoS20,T:SAT06,BSB:SAT15,SRSM:IJCAI19}.

\autoref{table:exactoverview} provides an overview on the tools we identified.
In a report of the model counting competition 2020 (MC2020), Fichte~\etal~\cite{FHH:JEA21} list existing tools capable of computing the number of satisfying assignments including d-DNNF compilers. 
In our empirical evaluation, we consider each d-DNNF compiler appearing in their list, namely \ctod{}~\cite{D:AAAI02}, \dsharp{}~\cite{MMBH:AAAI10}, and \dfour{}~\cite{LM:IJCAI17}.
For d-DNNF reasoners, no tool capable of reusing d-DNNFs for repetitive queries have been empirically evaluated or considered in another publication, to the best of our knowledge.
However, we extended our search to tools without scientific publication and found two providing such functionality, namely the reasoner \queryddnnf{} developed by Caridriot~\etal\footnote{\url{https://www.cril.univ-artois.fr/KC/d-DNNF-reasoner.html}} and a new version of \dfour{} on GitHub.
The extended (compared to the original work~\cite{LM:IJCAI17}) version of \dfour{}, does not allow reasoning on an existing d-DNNF, but allows to compute the number of satisfying assignments for several queries for a given CNF which we also include in our evaluation (referred to as \dfourquery{}).
We evaluate our proposal \ddknnife{}, \queryddnnf{} and \dfourquery{} as d-DNNF-based reasoners.
In preliminary experiments, we found that using \queryddnnf{} in combination with \dsharp{} always results in cardinalities of zero which may be caused by \dsharp{} violating the standard specified by \ctod{} in some cases.
Thus, we exclude the combination of \queryddnnf{} and \dsharp{}.
\queryddnnf{} does not support the output format of the latest \dfour{} version.
Therefore, we evaluate \queryddnnf{} with an earlier version.\footnote{\url{https://www.cril.univ-artois.fr/KC/d4.html}}

As baseline for evaluating our approach using d-DNNFs, we include the \ssat{} solvers \sharpsat{}, \countantom{}, and \ganak{}.
\doubleblind{In recent work~\cite{SHN+:EMSE23}, we identified those three solvers as the fastest \ssat{} for analyzing feature models.}{In recent work~\cite{SHN+:EMSE23}, those three solvers have been found very efficient for analyzing feature models compared to other \ssat{} solvers.}
Also, \ganak{} won the model counting competition 2020~\cite{FHH:JEA21}, while \sharpsat{} won in 2021 and 2022.\footnote{\url{https://mccompetition.org/index.html}}

Only \countantom{} and \ddknnife{} support multi-threading. For better comparability with the other tools, we decided to evaluate every tool with a single thread. As each considered d-DNNF compiler and \ssat{} solver requires an CNF as input format, we translate every feature model to CNF using FeatureIDE~\cite{MTS+17} prior to the experiments.

\begin{table}[h]
	\caption{Overview Evaluated Tools}
	\label{table:exactoverview}
	\centering
	\begin{tabular}{  l l l }
		\toprule
		Solver        & Type            & Reference                 \\
		\midrule
		\countantom{} & \ssat{} Solver  & \cite{BSB:SAT15}          \\
		\ganak{}      & \ssat{} Solver  & \cite{SRSM:IJCAI19}       \\
		\sharpsat{}   & \ssat{} Solver  & \cite{T:SAT06}            \\ \addlinespace[2pt]
		\ctod{}       & d-DNNF Compiler & \cite{D:AAAI02,D:ECAI04}  \\
		\dfour{}      & d-DNNF Compiler & \cite{LM:IJCAI17}         \\
		\dsharp{}     & d-DNNF Compiler & \cite{MMBH:AAAI10}        \\ \addlinespace[2pt]
		\ddknnife{}   & d-DNNF Reasoner & This Submission           \\
		\queryddnnf{} & d-DNNF Reasoner & No scientific publication \\
		\dfourquery{} & d-DNNF Reasoner & No scientific publication \\
		\bottomrule
	\end{tabular}

\end{table}

\paragraph{Experiment 1: Compilation to d-DNNF} In the first experiment, we measure the runtimes required to translate CNFs representing our subject systems to d-DNNF. Hereby, we evaluate each d-DNNF compiler, namely \ctod{}, \dsharp{}, and \dfour{}, on all 62 subject systems with a timeout of 30 minutes per feature model.
To put the runtime required for compilation into perspective, we also evaluate the runtime of computing the number of valid configurations using the three \ssat{} solvers, namely \sharpsat{}, \countantom{}, and \ganak{}.
According to insights from preliminary experiments, we set a timeout of five minutes.
Further, a \ssat{} solver exceeding the timeout of five minutes is very unlikely to scale for the repetitive queries in Experiment 2 and 3.
For each combination of compiler and subject system, we perform 50 repetitions. The compiled d-DNNFs are used for Experiment 2 and Experiment 3. The insights from Experiment 1 are used to answer RQ1.

\paragraph{Experiment 2: Cardinality of Features} In the second experiment, we measure the runtimes required to compute the cardinalities of all features in a given feature model. Hereby, we evaluate each d-DNNF tool combination (consisting of compiler and reasoner) and \ssat{} solver on each subject system. We set a timeout of 30 minutes for computing the cardinality of all features for a single feature model. For each measurement, we perform 50 repetitions. The insights from Experiment 2 are used to answer RQ2--RQ4.

\paragraph{Experiment 3: Cardinality of Partial Configurations} In the third experiment, we measure the runtimes required to compute the cardinality of partial configurations for a given feature model.
We randomly generate 250 partial configurations for each subject system.
We ensure that each partial configuration is satisfiable, as we assume that satisfiable configurations better reflect practical usage of feature-model counting based on the applications found in the literature~\cite{SNB+:VaMoS21}.
To generate the partial configurations, we iteratively select or deselect a randomly chosen feature.
If this results in an invalid configuration, we discard the current selection and continue.
The 250 configurations are separated in 50 chunks with sizes of 2, 5, 10, 20, and 50 features (randomly included or excluded), respectively.
Eleven feature models contain fewer than 50 features, in this case we only generate the configuration sets that contain fewer features than the respective feature model.
With the different configuration sizes, we aim to cover a wide range of analyses dependent on the cardinality of partial configurations~\cite{SNB+:VaMoS21}.
As for Experiment 2, we evaluate each d-DNNF tool combination and \ssat{} solver on each subject system. We set a timeout of 30 minutes for computing the cardinality of all 250 partial configurations for a single feature model. For each measurement, we perform 50 repetitions. The insights from Experiment~3 are used to answer RQ2--RQ4.

\paragraph{Experiment 4: Optimizations for ddnnife}
In the fourth experiment, we examine the impact of the optimizations presented in \autoref{sec:concept:optimizations}.
In particular, we compare different variants of \ddknnife{} with varying optimizations enabled.
\autoref{table:evaluation:optimizations} shows the list of \ddknnife{} variants we consider for evaluation.
First, we consider a naive approach without any optimizations that recursively traverses the whole d-DNNF without reusing any subtrees for each query.
The second variant also recursively traverses the d-DNNF for each query but reuses subtrees.
For the third, fourth, and fifth variant, we disable one optimization each: partial traversals, partial calculatios, and core/dead features, respectively.
As baseline we use the fully optimized version of \ddknnife{} used in prior experiments as sixth variant.
For each of the \ddknnife{} variants, we repeat the experiment design of Experiment~2 and Experiment~3 to measure the impact of the optimizations on computing the cardinality of features and partial configurations.
Hereby, we use the same input d-DNNFs, input partial configurations, and timeouts.
In addition to the repeated experiments, we evaluate the optimizations on unsatisfiable partial configurations as some optimizations cannot really be applied to unsatisfiable partial configurations.
The insights from Experiment~4 are used to answer RQ4.

\begin{table}
	\caption{\ddknnife{} Optimization Variants}
	\label{table:evaluation:optimizations}
	\centering\begin{tabular}{l l}
		\toprule
		Name                   & Enabled Optimizations               \\
		\midrule
		Naive                  & None                                \\
		Reusing Subtrees       & Reuse of subtrees                   \\
		No Partial Traversal   & All but partial traversals          \\
		No Partial Calculation & All but partial calculations        \\
		No Core/Dead           & All but usage of core/dead features \\
		ddnnife                & All                                 \\
		\bottomrule
	\end{tabular}

\end{table}

\paragraph{Statistical Tests}
We use a Mann-Whitney significance test~\cite{MN:CEP2010} to compare the performance of two tools for a feature model based on the 50 repetitions, as the samples are unpaired and we do not assume a normal distribution.
Hereby, we always assume a significance level of $\alpha = 5\%$.
To examine the effect sizes for significant results, we use Cohen's d~\cite{SF:JGME12}.
For rating the effect size, we use the classification shown in \autoref{table:experimentdesign:cohen} as suggested by Sullivan and Feinn~\cite{SF:JGME12}.
For instance, a value of $0.5 \leq d < 0.8$ as medium effect size.

\begin{table}[htbp]
	\centering\begin{tabular}{l r}
		\toprule
		Cohen's $d$        & Effect Size Classification \\
		\midrule
		$0.0 \leq d < 0.2$ & Very Small                 \\
		$0.2 \leq d < 0.5$ & Small                      \\
		$0.5 \leq d < 0.8$ & Medium                     \\
		$0.8 \leq d < 1.3$ & Large                      \\
		$1.3 \leq d$       & Very Large                 \\

		\bottomrule
	\end{tabular}
	\caption{Classification Effect Sizes of Cohen's d Adapted from Sullivan and Feinn~\cite{SF:JGME12}}
	\label{table:experimentdesign:cohen}
\end{table}

\paragraph{Technical Setup}
The entire empirical evaluation was performed on an Ubuntu 20.04.3 LTS server with a 64-bit architecture. The processor is an Intel(R) Xeon(R) CPU E5-260v3 with 2.40Ghz clock rate. Overall, 256 GB of RAM were available. During the measurements, the machine was only used for our evaluation to reduce the impact of other processes.
For each tool, the memory limit is set to 8 GB as we consider 8 GB to be a reasonable limit for hardware (\eg, notebooks or PCs) in practice.
Furthermore, in preliminary experiments we found no major differences in measured runtimes if we further increased the memory limit.
To measure runtimes, we used a Python-based framework which internally uses the module \texttt{timeit}.\footnote{\url{https://docs.python.org/3/library/timeit.html}}

\subsection{Results} \label{subsec:eval:results}
\paragraph{Experiment 1}
\begin{figure}
	\centering
	\begin{tikzpicture}
		\begin{axis}[
				width= \columnwidth,
				height = 0.6\columnwidth,
				ymode=log,
				ylabel=Runtime in Seconds (log),
				ylabel near ticks,
				xlabel=Subject System (Sorted by \#Features),
				xlabel near ticks,
				legend style={
						at={(0.5,-0.25)},anchor=north
					},
				legend columns=5,
				xmin=-1,
				xmax=63
			]
			\addplot[pointplot,mark=oplus,ctod,error bars/.cd, y dir = both, y explicit] table [x=Index, y=C2d-avg,  y error=C2d-stdev, col sep=comma] {\expone};
			\addplot[dfour,pointplot,mark=otimes,error bars/.cd, y dir = both, y explicit] table [x=Index, y=D4new-avg,  y error=D4new-stdev, col sep=comma] {\expone};
			\addplot[dfourold,pointplot,mark=otimes,error bars/.cd, y dir = both, y explicit] table [x=Index, y=D4-avg,  y error=D4-stdev, col sep=comma] {\expone};
			\addplot[dsharp,pointplot,mark=o,error bars/.cd, y dir = both, y explicit] table [x=Index, y=Dsharp-avg,  y error=Dsharp-stdev, col sep=comma] {\expone};
			\addplot[mark=none, red,domain=0:62] {1800};
			\legend{c2d, d4, d4-old, dSharp, Timeout/Error}
		\end{axis}
	\end{tikzpicture}
	\caption{Runtime for Offline Phase of d-DNNF Compilers (Average \& Standard Deviation)}
	\label{dia:exp1:compilation}
\end{figure}
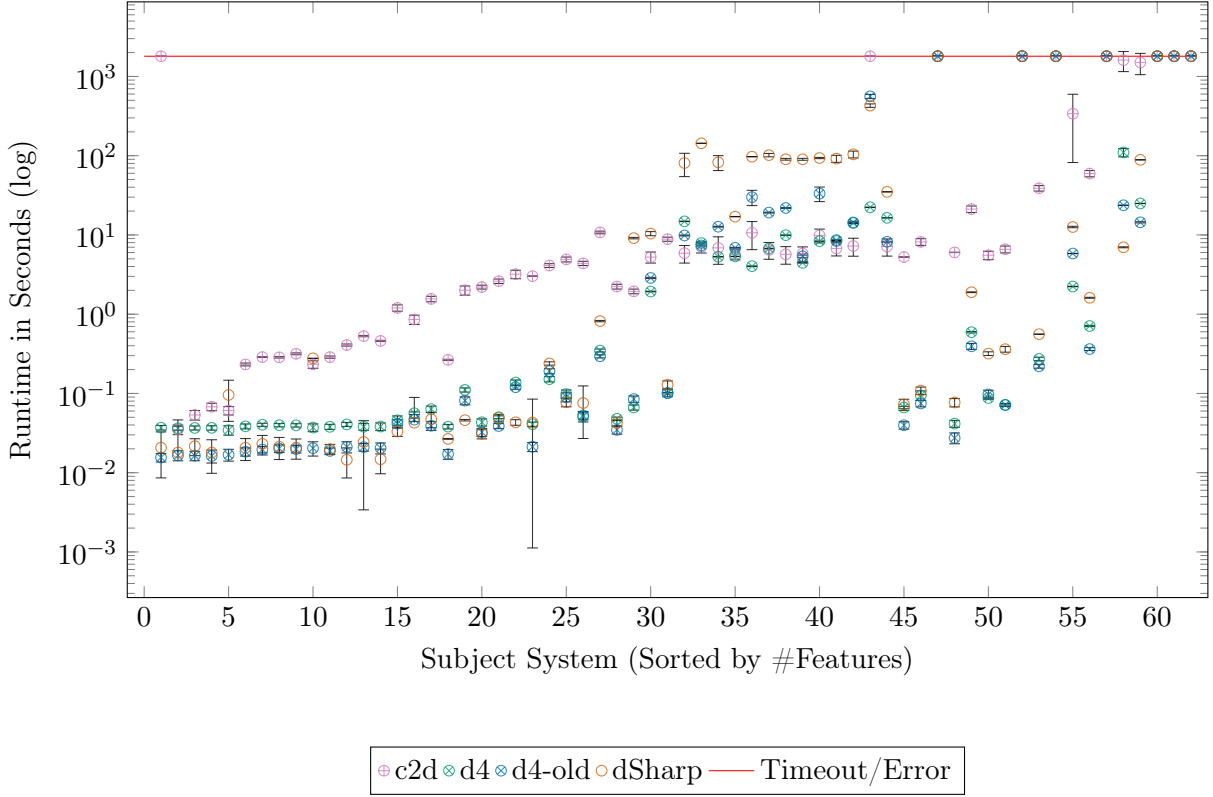

\autoref{dia:exp1:compilation} shows the runtime of the d-DNNF compilers for transforming each subject system.
Each point on the x-axis corresponds to a subject system.
The y-axis shows the runtime average and standard deviation for the 50 repetitions of the compilers with a logarithmic scale.
For 55 out of the 62 subject systems, at least one d-DNNF compiler successfully compiled the CNF into d-DNNF within 30 minutes.
The three \ssat{} solvers hit the timeout of five minutes (\cf \autoref{subsec:eval:experimentdesign}) for the same seven feature models.\footnote{Four Linux variants, Embtoolkit, Freetz, and Automotive05.}
52 subject systems were compiled within 10 seconds by at least one of the compilers.
\dfour{} (both versions) and \dsharp{} successfully compiled 55 feature models while \ctod{} successfully compiled 53.
For LLVM, a subject system consisting of a single unit clause, \ctod{} throws an error which is probably caused by preprocessing on unit clauses that leave an illegal (empty) formula.
For the 53 models successfully evaluated by each compiler, \dfour{}, \dsharp{}, and \ctod{} on average required 4.62, 21.5, and 69.5 seconds, respectively.
The majority (83\%) of d-DNNFs require less than 10 MB of space.
The d-DNNFs produced by the fastest compiler (\dfour{}) require at most 20.1 MB (median 0.04 MB).
% \dfour{}, \dsharp{}, and \ctod{} are significantly faster ($p < 9\cdot10^{-5}$) than the other compilers for 21, 9, and 11 feature models, respectively.%

\paragraph{Experiment 2}
All tools except \dsharp{} and \sharpsat{} computed the same results (\ie, cardinalities) for each feature of all considered feature models.
The results of \dsharp{} (as black box) and \sharpsat{} differed from those results for 55.28\% (31 affected models) and 15.32\% (31) of the overall computed cardinalities, respectively.
\ddknnife{} with d-DNNFs compiled by \dsharp{} differed for 0.54\%.

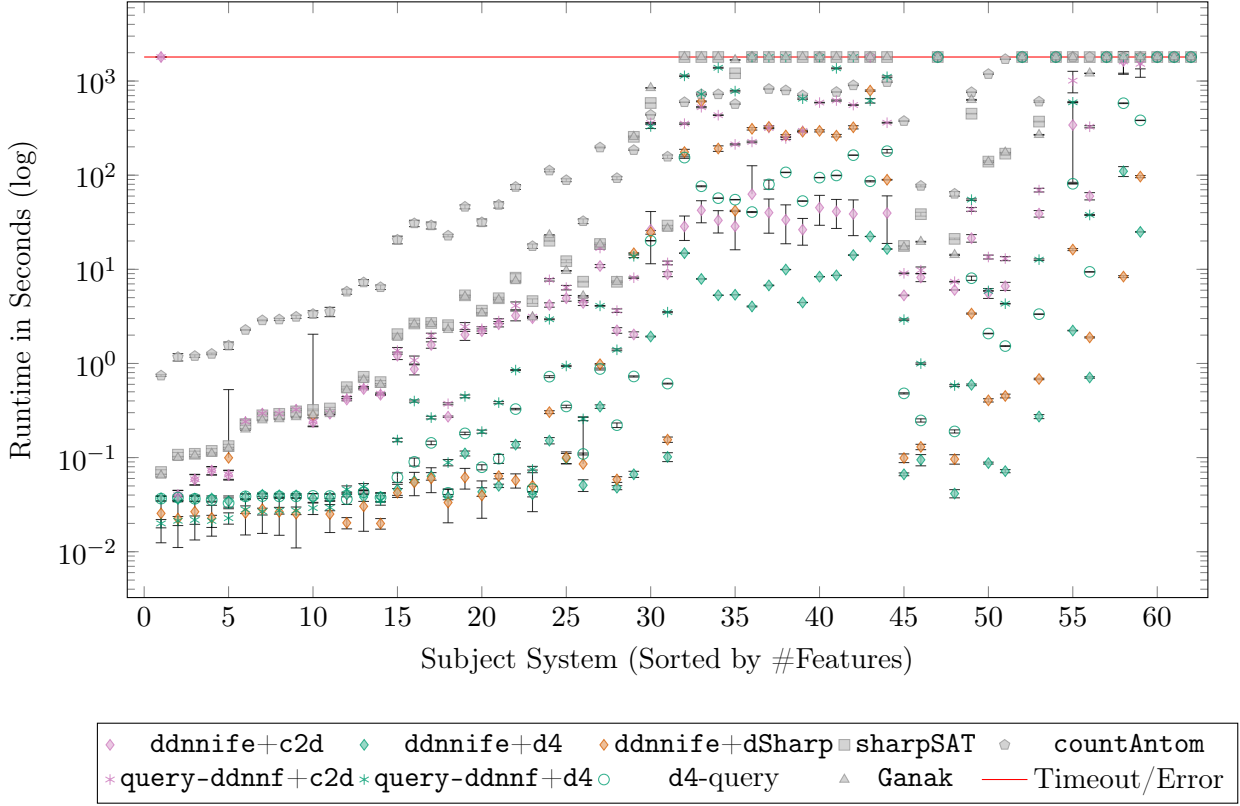
\begin{figure}
	\centering
	\begin{tikzpicture}
		\begin{axis}[
				ymode=log,
				width= \columnwidth,
				height = 0.6\columnwidth,
				ylabel=Runtime in Seconds (log),
				ylabel near ticks,
				xlabel=Subject System (Sorted by \#Features),
				xlabel near ticks,
				legend style={
						at={(0.5,-0.21)},anchor=north
					},
				legend columns=5,
				xmin=-1,
				xmax=63
			]
			\addplot[ddknnife,ctod,error bars/.cd, y dir = both, y explicit] table [x=Index, y=Ddknnifec2d-avg,  y error=Ddknnifec2d-stdev, col sep=comma] {\exptwo};
			\addplot[ddknnife,dfour,error bars/.cd, y dir = both, y explicit] table [x=Index, y=Ddnnifed4-avg,y error=Ddnnifed4-stdev, col sep=comma] {\exptwo};
			\addplot[ddknnife,dsharp,error bars/.cd, y dir = both, y explicit] table [x=Index, y=Ddknnifedsharp-avg,y error=Ddknnifedsharp-stdev, col sep=comma] {\exptwo};
			\addplot[sharpsat,error bars/.cd, y dir = both, y explicit] table [x=Index, y=Sharpsat-avg,  y error=Sharpsat-stdev, col sep=comma] {\exptwo};
			\addplot[countantom,error bars/.cd, y dir = both, y explicit] table [x=Index, y=Countantom-avg,y error=Countantom-stdev, col sep=comma] {\exptwo};
			\addplot[queryddnnf,ctod,error bars/.cd, y dir = both, y explicit] table [x=Index, y=Queryc2d-avg,  y error=Queryc2d-stdev, col sep=comma] {\exptwo};
			\addplot[queryddnnf,dfour,error bars/.cd, y dir = both, y explicit] table [x=Index, y=Queryd4-avg,y error=Queryd4-stdev, col sep=comma] {\exptwo};
			\addplot[dfourquery,error bars/.cd, y dir = both, y explicit] table [x=Index, y=D4query-avg,y error=D4query-stdev, col sep=comma] {\exptwo};

			\addplot[ganak,error bars/.cd, y dir = both, y explicit] table [x=Index, y=Ganak-avg,y error=Ganak-stdev, col sep=comma] {\exptwo};
			\addplot[mark=none, red,domain=0:62] {1800};
			\legend{\ddknnife{}+\ctod{},\ddknnife{}+\dfour{}, \ddknnife{}+\dsharp{},  \sharpsat{}, \countantom{}, \queryddnnf{}+\ctod{}, \queryddnnf{}+\dfour{}, \dfour{}-query,  \ganak{}, Timeout/Error}
		\end{axis}
	\end{tikzpicture}
	\caption{Runtime for Offline + Online Phase of Cardinality of Features (Average \& Standard Deviation)}
	\label{dia:exp2}
\end{figure}

\autoref{dia:exp2} shows the runtimes required to compute the cardinality of all features using d-DNNFs (considering offline and online phase) and repetitive \ssat{} calls.
For each feature model successfully compiled to d-DNNF (55 of 62), at least one d-DNNF tool combination was able to successfully compute all feature cardinalities  within 30 minutes.
The \ssat{} solvers evaluated all feature cardinalities for only 49 feature models.
For every feature model, \ddknnife{} combined with \dfour{} is significantly faster than every \ssat{} solver ($p < 7.1 \cdot 10^{-18}$) with very large effect sizes ($d \geq 8.5$).

\autoref{table:exp2:sharessat} shows the relative share of features that the three \ssat{} solvers were able to evaluate within the 30 minutes timeout.
While the fastest d-DNNF tool combination (\ddknnife{} and \dfour{}) for \textit{Embtoolkit} required 48.3 seconds to evaluate its feature cardinalities, the fastest \ssat{} solver (\sharpsat{}) required 30 minutes to evaluate 0.86\% of the features.
Extrapolating the runtimes, \sharpsat{} would require 58 hours to evaluate all features for \textit{Embtoolkit} which is 8,316 (39,253 without compilation) times slower than \ddknnife{} with \dfour{}.
Also considering such extrapolations, the fastest d-DNNF-based tool combination is on average 480 times faster than the fastest \ssat{} solver.
For the feature models with more than 2,000 features, the fastest d-DNNF-based tool combination is on average even 2,307 times faster.
While all 55 feature models were evaluated within 2.8 minutes with d-DNNFs, \ssat{} would require 67 \textit{hours} (1,421 times slower), considering the fastest tool per instance.
Considering the compilation time and average computation time to evaluate a feature, the fastest combination of d-DNNF compiler and reasoner breaks even after 3.9 queries (median) with the fastest \ssat{} solver.
For every feature model that requires more than 0.1 seconds to be evaluated, \ddknnife{} with \dfour{} is significantly faster ($p < 7.1\cdot 10^{-18}$) with very large effect sizes ($d \geq 4.86$) than any tool or tool combination without \ddknnife{}.
For the online phase, considering the fastest combination with a compiler for both reasoners, \ddknnife{} requires 4.3 minutes for all feature models while \queryddnnf{} requires 72 minutes.
Using the partial traversals (\cf \autoref{sec:concept:optimizations}), \ddknnife{} needs to only recompute 6.5\% (compiled by \dfour{}), 4.4\% (\dsharp{}), and 10.1\% (\ctod{}) of the d-DNNF nodes per feature on average.

\begin{table}[b]
	\caption{Share of Evaluated Features for \ssat{} Solvers}
	\label{table:exp2:sharessat}
	\centering
	\begin{tabular}{@{}lrrr@{}}
		\toprule
		System           & CountAntom & Ganak  & SharpSAT \\ \midrule
		se77x            & 98.1\%     & 21.3\% & 47.23\%  \\
		integrator\_arm9 & 91.7\%     & 23.4\% & 31.3\%   \\
		Freebsd          & 46.1\%     & 4.87\% & 5.73\%   \\
		Automotive07a    & 5.47\%     & 30.2\% & 36.1\%   \\
		Automotive02     & 2.28\%     & 17.5\% & 19.6\%   \\
		Embtoolkit       & 0.09\%     & 0.65\% & 0.86\%   \\ \bottomrule
	\end{tabular}
\end{table}

\paragraph{Experiment 3}
All tools but \dsharp{} computed the same cardinalities for every partial configuration and feature model.
The results of \dsharp{} (black box) and \ddknnife{} with \dsharp{} differed from those results in 41.83\% (31 affected models) and 5.83\% (31 affected models) cases, respectively.

\begin{figure}
	\centering
	\begin{tikzpicture}
		\begin{axis}[
				width= \columnwidth,
				height = 0.6\columnwidth,
				ymode=log,
				ylabel=Runtime in Seconds (log),
				ylabel near ticks,
				xlabel=Subject System (Sorted by \#Features),
				xlabel near ticks,
				legend style={
						at={(0.4,-0.21)},anchor=north
					},
				legend columns=5,
				xmin=-1,
				xmax=63
			]
			\addplot[ddknnife,ctod,error bars/.cd, y dir = both, y explicit] table [x=Index, y=Ddknnifec2d-avg,  y error=Ddknnifec2d-stdev, col sep=comma] {\expthree};
			\addplot[ddknnife,dfour,error bars/.cd, y dir = both, y explicit] table [x=Index, y=Ddnnifed4-avg,y error=Ddnnifed4-stdev, col sep=comma] {\expthree};
			\addplot[ddknnife,dsharp,error bars/.cd, y dir = both, y explicit] table [x=Index, y=Ddknnifedsharp-avg,y error=Ddknnifedsharp-stdev, col sep=comma] {\expthree};
			\addplot[sharpsat,error bars/.cd, y dir = both, y explicit] table [x=Index, y=Sharpsat-avg,  y error=Sharpsat-stdev, col sep=comma] {\expthree};
			\addplot[countantom,error bars/.cd, y dir = both, y explicit] table [x=Index, y=Countantom-avg,y error=Countantom-stdev, col sep=comma] {\expthree};
			\addplot[queryddnnf,ctod,error bars/.cd, y dir = both, y explicit] table [x=Index, y=Queryc2d-avg,  y error=Queryc2d-stdev, col sep=comma] {\expthree};
			\addplot[queryddnnf,dfour,error bars/.cd, y dir = both, y explicit] table [x=Index, y=Queryd4-avg,y error=Queryd4-stdev, col sep=comma] {\expthree};
			\addplot[dfourquery,error bars/.cd, y dir = both, y explicit] table [x=Index, y=D4query-avg,y error=D4query-stdev, col sep=comma] {\expthree};

			\addplot[ganak,error bars/.cd, y dir = both, y explicit] table [x=Index, y=Ganak-avg,y error=Ganak-stdev, col sep=comma] {\expthree};
			\addplot[mark=none, red,domain=0:62] {1800};
			\legend{\ddknnife{}+\ctod{},\ddknnife{}+\dfour{}, \ddknnife{}+\dsharp{},  \sharpsat{}, \countantom{}, \queryddnnf{}+\ctod{}, \queryddnnf{}+\dfour{}, \dfour{}-query,  \ganak{}, Timeout/Error}
		\end{axis}
	\end{tikzpicture}
	\caption{Runtime for Offline + Online Phase of Cardinality of Partial Configurations (Average \& Standard Deviation)}
	\label{dia:exp3}
\end{figure}
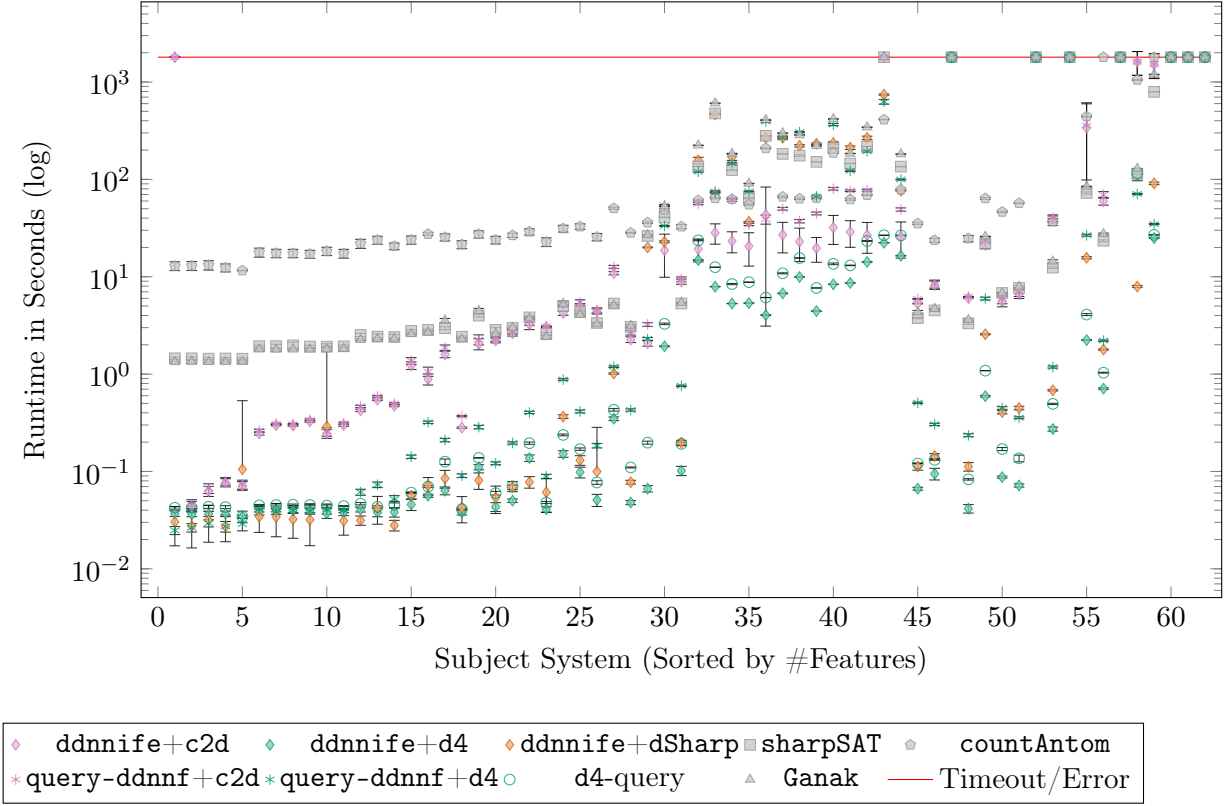

\autoref{dia:exp3} shows the runtimes required to compute the cardinality of the 250 generated partial configurations with d-DNNFs and repetitive \ssat{} calls.
\ddknnife{} and \queryddnnf{} in combination with any compiler and \dfourquery{} are able to compute the cardinality for the 250 partial configurations within 30 minutes for every successfully compiled d-DNNF (overall 55).
At least one \ssat{} solver successfully evaluated the partial configurations for 53 feature models.
For all successfully evaluated feature models, \ddknnife{} with \dfour{} is significantly faster (p < $7.1 \cdot 10^{-18}$) than the fastest \ssat{} solver with very large effect sizes ($d \geq 8.6$).
For every feature model, the fastest d-DNNF-based tool combination is between 4.1 and 394 times faster than the fastest \ssat{} solver.
Overall, the fastest runtimes amount to 3.7 minutes for d-DNNF and 44 minutes for \ssat{} (11.9 times slower).
The fastest tool combination with \ddknnife{} requires 5.23 minutes to evaluate all instances while \dfourquery{} and combinations with \queryddnnf{} require 5.88 and 24 minutes, respectively.

\begin{figure}
	\centering
	\includegraphics[width=\textwidth]{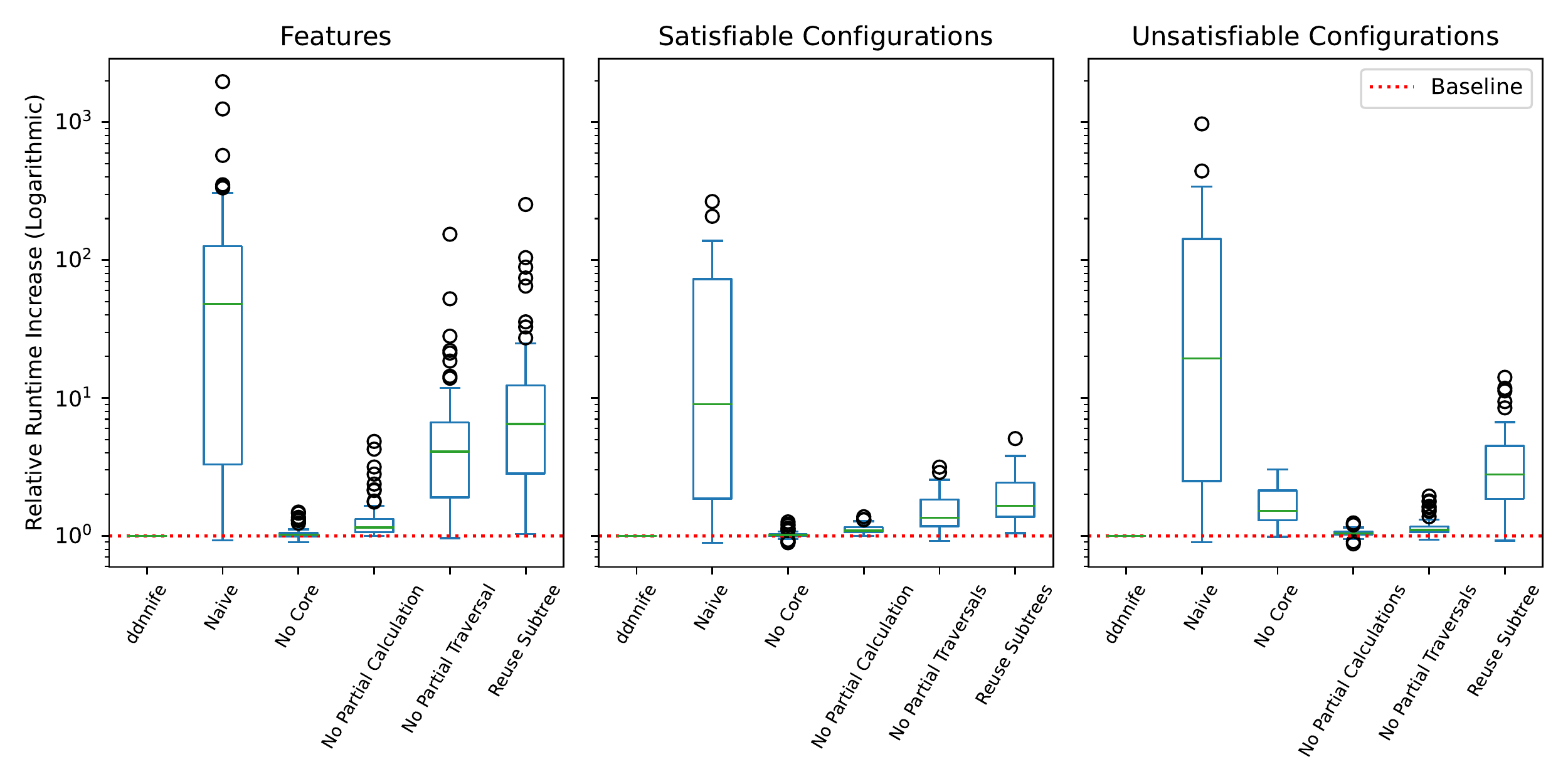}
	\caption{Runtime \ddknnife{} Variants on \dfour{} d-DNNFs}
	\label{dia:exp4}
\end{figure}

\paragraph{Experiment 4}
For Experiment~4, we only show plots for the the runtimes of \ddknnife{} for d-DNNFs compiled by \dfour{} as the results for d-DNNFs from \dsharp{} and \ctod{} are highly similar.
Nevertheless, the results for \ctod{} and \dsharp{} are included in our replication package.
\autoref{dia:exp4} shows the relative increase in runtime of the different \ddknnife{} variants compared to the fully optimized variant.
For instance, a value of 1,000 indicates that a \ddknnife{} was 1,000 times slower than the fully optimized variant for the given feature model.
The plots are separated in the runtimes required for cardinality of features (left), cardinality of satisfiable configurations (middle), and cardinality of unsatisfiable configurations (right).

For features, \ddknnife{} with all optimizations requires the least runtime with 5.29 minutes for all 55 feature models.
Disabling the optimizations for core/dead features and partial calculations, results in similar runtimes with 5.31 and 6.40 minutes of runtime, respectively.
Without partial traversals, \ddknnife{} requires 30.0 minutes to evaluate the 55 feature models which is 5.67 times slower than the fully optimized version.
The fully naive variant, not using any optimizations presented in \autoref{sec:concept:optimizations}, hits the timeout of 30 minutes for 13 of 55 feature models.
None of the other variants hits the timeout once.
For each feature model that requires more than 0.1 seconds to be evaluated the fully optimized \ddknnife{} variant is significantly faster ($p < 0.025$) than any other variant but the variant with \textit{no core}.
Not using core/dead features as optimization is significantly faster ($p < 6.5 \cdot 10^{-7}$) than the fully optimized variant for 12 feature models.

For satisfiable partial configurations, the fully optimized variant and the variant without using core/dead features perform very similar with 4.24 and 4.26 minutes, respectively.
Further, the variants without partial calculations and partial traversals follow closely with 4.52 and 4.57 minutes, respectively.
The naive variant hits the timeout for 9 feature models.
The fully optimized variant is significantly faster ($p < 1.5 \cdot 10^{-17}$) than the naive variant and the variant without reusing subtrees for every feature model that requires more than 0.1 seconds to be evaluated.
For no partial calculation and partial traversals, the runtimes for one and two feature models do not differ significantly, respectively.
The variant without using core/dead features is significantly faster ($p < 0.029$) than \ddknnife{} with all optimizations for seven feature models. 

For unsatisfiable partial configurations, the fully optimized variant has the least overall runtime with 2.40 minutes.
The runtimes for leaving out partial calculations and partial traversals are very similar with an overall runtime of 2.51 and 2.60 minutes, respectively.
Without using the optimization based on core/dead features, \ddknnife{} requires 3.85 minutes which is 1.59 times slower than the fully optimized variant.
Further, the fully optimized variant is significantly faster ($ p < 2.0 \cdot 10^{-6}$) than the variant without core features for all but two feature models.
The relative increase in runtime without optimizing using core/dead features is much higher than for satisfiable configurations (1.01 times slower) and features (1.003 times slower).
\ddknnife{} without any optimizations hits the timeout of 30 minutes for 9 feature models.

\subsection{Discussion} \label{subsec:eval:discussion}

\noindent\textbf{RQ1} \textit{\rqone{}}
Our results indicate that compilation to d-DNNF is promising for feature-model counting.
90\% of the evaluated industrial feature models can be compiled within 30 minutes and 84\% even within 10 seconds.
\ssat{} solvers and d-DNNF compilers scaled to the exact same feature models.
Compiling to d-DNNF seems to only fail for feature models that are very hard for feature-model counting in general.
This observation matches the expectation as popular d-DNNF compilers and \ssat{} behave very similar~\cite{MMBH:AAAI10,T:SAT06,LM:IJCAI17,D:AAAI02}.
In most environments, the memory usage of d-DNNFs should also cause no problems, as, \eg, the fastest compiler \dfour{} produces d-DNNFs requiring at most 20.1 MB.
Overall, considering d-DNNF size and compilation time, \dfour{} is the most promising d-DNNF compiler for feature models.

\noindent\textbf{RQ2} \textit{\rqtwo{}}
\countantom{}, \ganak{}, \dfour{}, \ctod{}, \ddknnife{} (with \ctod{} and \dfour{}), and \queryddnnf{} computed the same cardinalities for every single computation which we consider as a strong indicator for the correctness of these tools.
In particular, due to more than 200k correct results, it is very likely that the algorithms and optimizations presented in this submission and implemented in \ddknnife{} yield correct results.
Only on faulty d-DNNFs produced by \dsharp{}, \ddknnife{} computes few differing cardinalities, which is very likely caused by \dsharp{} providing incorrect d-DNNFs in those cases.

\sharpsat{} and \dsharp{}, which is based on \sharpsat{}, compute different results from the remaining tools in many cases (\sharpsat{}: 13.4\%, \dsharp{}: 53.6\%). The differences may be related to an error in \sharpsat{} caused by unit clauses as discussed in other work~\cite{FHR:SAT22} and in a GitHub issue.\footnote{\url{https://github.com/marcthurley/sharpSAT/issues/5}}
While a potential fix could impact runtimes, we assume that our measurements may still be a reasonable indicator for the performance of the correct behavior due to the nature of the bug~\cite{FHR:SAT22}.
Further, as both tools perform similar to other included tools, we expect no substantial changes in the conclusions once the bug is fixed.
We consider the inclusion of the solvers for comparison as important as both tools have been employed in the product-line domain~\cite{OGB+:TR20,STS:VaMoS20,MOP+:SPLC19,LGCR:SPLC15,OGB:SPLC19}.
Therefore, we include \sharpsat{} and \dsharp{}, despite the bug, for the following research questions.

\noindent\textbf{RQ3} \textit{\rqthree{}}
Reusing d-DNNFs for repetitive queries of feature-model counting is significantly faster than using state-of-the-art \ssat{} solvers for every considered feature model.
For analyzing all 55 successfully evaluated feature models, using the fastest d-DNNF-based approach requires overall 6.56 minutes while the fastest \ssat{} solver would require 2.81 \textit{days} (619 times slower).
\ddknnife{} with \dfour{} is up-to 8,316 times faster than the fastest \ssat{} solver even including compilation.
Reusing the d-DNNF over both query types, and potentially even further analyses, would even further reduce the runtimes for d-DNNFs.
Overall, we argue the effort of compilation to d-DNNF is worth for feature-model counting and substantially more efficient than the current state-of-the-art of repetitive \ssat{} queries.
As d-DNNFs for repetitive counting queries have not been empirically evaluated in any domain before, our results provide the first insights on the performance.
The significant and substantial improvements over repetitive invoking \ssat{} solvers suggest that reusing d-DNNFs may also be promising for other domains.

Overall, the combination of \ddknnife{} and \dfour{} is faster than any other tool or tool combination.
For features, a combination of \ddknnife{} and \dfour{} is the fastest requiring 10 minutes for all feature models followed by \dfourquery{} (40 minutes) and \ddknnife{} with \dsharp{} (70).
For partial configurations, there are fewer queries and \ddknnife{} needs to traverse a larger part of the d-DNNF with its partial traversals (\cf \autoref{sec:concept:optimizations}) and, thus, its performance advantages are smaller.
While tool combinations including \ddknnife{} are still fastest for partial configurations overall, \dfourquery{} is faster for many feature models, especially smaller ones.
However, \dfourquery{} still comes with the disadvantage of requiring to start from scratch while \ddknnife{} and \queryddnnf{} may reuse d-DNNFs for further analyses and also allow processing d-DNNFs from multiple compilers.
Applying d-DNNFs with a combination of \dfour{} as compiler and \ddknnife{} as reasoner seems to be the most promising option for repetitive queries of feature-model counting.

\noindent\textbf{RQ4} \textit{\rqfour{}}
The presented optimizations are essential for the efficiency of \ddknnife{}.
The variant without any optimizations hits the timeout of 30 minutes for feature models which require a few seconds with the fully optimized variant.
In particular, partial traversals, iterative traversals, and reusing subtrees all substantially reduce the runtime.
Partial calculations and core/dead features have mostly marginal but still positive effects on the runtimes.
Using core/dead features for optimization only has a considerable effect for evaluating unsatisfiable partial configurations which matches the expectation.
For the cardinality of features, core and dead features require no traversal through the d-DNNF whatsoever, as none of the nodes is marked for a partial traversal.
Further, our approach of using core/dead features for partial configurations only prevents traversals for unsatisfiable configurations which reduces the benefits when considering only satisfiable configurations.

\subsection{Threats to Validity} \label{subsec:eval:threats}

\paragraph{Translating Feature Models to CNF}
Each evaluated compiler and \ssat{} solver uses CNF as input format.
To employ the tools, the feature models need to be translated to CNF.
Two equivalent but syntactically different CNFs may result in significantly varying runtime for solvers and compilers~\cite{KK07,OGB+:TR19}.
Hence, changing the CNF translation may have an impact on the measured runtimes.
However, analyzing multiple CNF translations is beyond the scope of this work.
Another aspect to consider is that some CNF translations are equisatisfiable but do not preserve the number of satisfying assignments.
Such a translation may lead to faulty cardinalities in our analyses.
For each feature model given in a proprietary format, we use the transformation to CNF employed by FeatureIDE~\cite{MTS+17} which presumably does not change the number of satisfying assignments.
The remaining feature models (collected by Oh~\etal~\cite{OGB+:TR20}) were already specified in CNF and have been used in other empirical evaluations~\cite{OGB+:TR20,OGB+:TR19,KGS+:ICSE19,PAP+:ICST19}.

\paragraph{Parameterization of Solvers/Compilers}
All evaluated \ssat{} solvers and compilers provide parameters that potentially impact their runtime.
It is possible that different parameters may significantly change the performance of evaluated tools.
However, we expect that using default parameters reasonably represents the usage in practice.
Parameter optimization for each solver and compiler is out of scope for this work.

\paragraph{Random Effects and Computational Bias}
When measuring the runtime of a solver required to analyze a feature model, the runtimes may vary between measurements due to computational bias, random effects, or background processes.
To reduce the impact of such influence factors, we perform 50 repetitions for each measurement.
Also, while performing measurements no other major processes were executed.
Furthermore, we employ established statistical tests to validate the significance of our conclusions.

\paragraph{Multi-Threaded Tools} In our empirical evaluation, we decided to evaluate all tools single threaded.
As our tool \ddknnife{} and the \ssat{} solver \countantom{} support multithreading, enabling it may improve the performance of both tools. However, the considered problems (\ie, cardinality of features and partial configurations) rely on multiple queries that could also be parallelized for other solvers.
Hence, we consider evaluating all tools singlethreaded as more sensible option.
Further, preliminary experiments showed that the performance of both tools only differs slightly when enabling multithreading.

\paragraph{External Validity of Experimental Setting} For a different experimental setting (\eg, fewer queries), the results and comparison between d-DNNF and \ssat{} may differ. However, the evaluated queries are motivated by the literature and our industry projects. For instance, analyses such as homogeneity by definition require to compute the cardinality of all features. Further, the d-DNNF-based approach was generally faster than repetitively invoking \ssat{} solvers even after a fraction of the queries.

\paragraph{External Validity of Tools}
The measured performance of the evaluated tools cannot be necessarily transferred to other tools.
However, to the best of our knowledge, we evaluated all publicly available d-DNNF compilers and reasoners.
For \ssat{} solvers, we selected the two best performing \ssat{} solvers for feature-model counting \doubleblind{according to our previous experiments~\cite{STS:VaMoS20}}{according to recent insights~\cite{STS:VaMoS20}} and the winning \ssat{} solver of the model counting competition 2020.
Hence, we expect our results to reasonably represent the performance of both approaches.

\paragraph{External Validity of Subject Systems}
Our conclusions do not necessarily hold for other real-world feature models.
However, we evaluated feature models from a variety of domains, namely  operating systems, automotive, financial services, and multiple other software system domains.
Furthermore, we covered a wide range of features (11--62,482), clauses (1--350,221), number of valid configurations (1,024--$10^{1534}$), and induced runtimes for solving \ssat{} (between few milliseconds and hitting timeouts of 24 hours~\cite{STS:VaMoS20}).

\section{Related Work} \label{sec:relatedwork}

In this section, we discuss work that (a) performs model counting with d-DNNFs, (b) uses d-DNNFs in the context of feature models, (c) proposes alternative approaches to allow feature-model counting, (d) applies knowledge compilation in feature-model analysis, or (e) presents other knowledge-compilation artifacts for which model counting is tractable.

\paragraph{Model Counting with d-DNNFs}
Each considered d-DNNF compiler, namely \ctod{}~\cite{D:AAAI02}, \dsharp{}~\cite{MMBH:AAAI10}, and \dfour{}~\cite{LM:IJCAI17}, is able to compute the number of satisfying assignments of the input. However, neither compiler allows reusing a given d-DNNF to compute cardinalities of features or partial configurations.
While \dfour{} (given a CNF) and \queryddnnf{} support computing multiple queries, those functionalities have not been considered in a scientific publication and have not been evaluated before. Generally, there has not been any empirical evaluation on reusing d-DNNFs for counting.

\paragraph{d-DNNF Exploitation in Feature-Model Analysis}
Sharma~\etal~\cite{SGRM:LPAIR18} use d-DNNFs for uniform random sampling with their tool \textit{KUS}.
Internally, \textit{KUS} depends on counting on a given d-DNNF.
While that counting procedure could be extracted, it does not support applying assumptions (here: including/excluding features) and, thus, cannot be used for both query types we consider.
Bourhis~\etal~\cite{BDD+:TR23} employ d-DNNFs for multiple different feature-model analyses including counting. However, for counting, they only consider single computations for the entire feature model while we target repetitive queries for features and partial configurations.
In previous publications, we~\cite{STS:VaMoS20} and Kübler~\etal~\cite{KZK:LoCoCo10} evaluate d-DNNF compilers for the task of computing the number of valid configurations. However, in both publications the compilers are just used as black-box \ssat{} solvers, essentially compiling a new d-DNNF for every single computation, and do not reuse the compiled d-DNNF. Furthermore, no repeated queries (\eg, for cardinality of features) are performed.

\paragraph{Solutions for Feature-Model Counting}
In the above mentioned works, Kübler~\etal~\cite{KZK:LoCoCo10} and we~\cite{STS:VaMoS20} compute the cardinalities of industrial feature models. Both works are limited to one query per feature model and only use the solvers as black-box, whereas we reuse the compiled the d-DNNFs to reduce the effort of repetitive queries.

Pohl~\etal~\cite{PLP:ASE11} compare the runtimes of several solver types, namely BDD, SAT, and CSP, on different feature-model analyses, including feature-model counting.
The results indicate that BDDs scale substantially better than SAT and CSP for computing the number of valid configurations.
However, their dataset contains only comparatively small feature models (less than 300), whereas our benchmark contains feature models with several ten-thousands features.
In previous work, we found that compiling to BDD scales for only very few industrial feature models and considerably worse than d-DNNFs~\cite{SHN+:EMSE23,HST:SPLC21}.

Other work proposes ad-hoc solutions for feature model counting targeting feature models with no~\cite{HFCC:IET11} or very few cross-tree constraints~\cite{FHCC:TSE14}.
While those perform well for the feature models with no or very few cross-tree constraints, recent work indicates that industrial feature models tend to have very large number of cross-tree constraints (\eg, up-to 15k in our dataset)~\cite{STS:VaMoS20,OGB+:TR19,HST:SPLC21, KTM+:SE18}.

\paragraph{Uniform-Random Sampling}
Several approaches for uniform random sampling internally depend on counting~\cite{SGRM:LPAIR18,MOP+:SPLC19,OGB+:TR20,AZT:SAT18}.
For instance, \textit{Smarch}~\cite{OGB+:TR19} builds a uniform sample by repetitively invoking \sharpsat{}. \textit{Spur}~\cite{AZT:SAT18} adapts \sharpsat{} to cache values of computation nodes to accelerate deriving valid configurations. However, like for \textit{KUS} (see \textit{d-DNNF Exploitation in Feature-Model Analysis}), it is not possible to apply assumptions for \textit{Spur}. Hence, computing the cardinality of features or partial configurations is also not possible without major conceptual adaptations.

\paragraph{Knowledge Compilation in Feature-Model Analysis}
Binary decision diagrams (BDDs) are often applied for feature-model analyses~\cite{ACLF:ASE11,AHC+:CAiSE12,BSTRC:VaMoS07,MWCC:GPCE08,STS:VaMoS20,OBMS16}, including some proposals to use BDDs for feature-model counting~\cite{HST:SPLC21,JHF+:ICTSS12,MBC:OOPSLA09,PLP:ASE11,STS:VaMoS20}.
However, BDDs fail to scale for many industrial feature models~\cite{HST:SPLC21,STS:VaMoS20,OGB+:TR19,T:SPLC20}.
Krieter~\etal~\cite{KTS+:ICSE18,KAN+:SPLC21} employ modal implication graphs to accelerate decision propagation during product configuration.
However, modal implication graphs are not tractable for feature-model counting.

\paragraph{Model Counting with Other Knowledge-Compilation Artifacts}
Model counting is tractable for several knowledge-compilation artifacts (\ie, has a polynomial time complexity with respect to the size of artifacts).
According to the knowledge compilation map of Darwiche and Marquis~\cite{DM:JAIR02}, binary decision diagrams, d-DNNFs, and MODs (a DNF that satisfies smoothness and determinism) allow polynomial-time counting. In other work, sentential decision diagrams~\cite{OD:IJCAI15} and affine decision trees~\cite{KLMT:IJCAI13} have been found tractable for model counting. While all these formats allow counting in theory, \doubleblind{we compared the performance of BDD, SDD, EADT, and d-DNNF compilers for the task of feature-model counting in recent work~\cite{SHN+:EMSE23} and found d-DNNF compilers to be the most efficient.}{d-DNNF compilers have been found the most efficient for compiling feature models, considering off-the-shelf tools~\cite{STS:VaMoS20}.}

\paragraph{Incremental Solving}
An incremental solver performs multiple queries on highly similar formulas while keeping information between those queries to accelerate later invocations~\cite{FBS:SAT19}.
For regular SAT, several solvers were proposed with promising results for many use-cases~\cite{FBS:SAT19,ES:ENTCS03,NRS:SAT14}.
Typically, those solvers are adaptations of popular SAT solvers based on DPLL~\cite{GHN+:CAV04} or CDCL~\cite{MLM09}.
While such incremental SAT solvers have been used for years, there has not been such adaptations for \ssat{}.

\section{Conclusion}
Many analyses based on feature-model counting require various similar counting queries to compute the number of valid configurations that contain a certain feature or conform to a partial configuration.
We are the first to reuse d-DNNFs to reduce the effort of repetitive counting queries.
Our evaluation shows that reusing d-DNNFs substantially reduces the runtime compared to the state-of-the-art of repetitively invoking \ssat{} solvers.
We found that compilation to d-DNNF scales for every feature model, for which state-of-the-art \ssat{} solvers are able to compute a result.
Further, for all 55 successfully evaluated feature models, our tool \ddknnife{}, combined with the fastest compiler, requires 6.5 \textit{minutes} to evaluate all cardinalities while the fastest \ssat{} solver requires 2.81 \textit{days} of runtime.
Our tool \ddknnife{} is significantly faster than using the fastest \ssat{} solver for every considered feature model, saving up-to 99.99\% of CPU time even when including the compilation time.
If we exclude the compilation time, \ddknnife{} is even up-to 39,253 times faster than the fastest \ssat{} solver.
Overall, we recommend reusing d-DNNFs with \ddknnife{} for feature-model counting when multiple queries are required.

\section*{Acknowledgments}
This work bases on the theses of Raab~\cite{Raab22} and Sundermann~\cite{Sundermann20}.
	
\printbibliography

\end{document}